\documentclass[reprint,amsmath,amssymb,
superscriptaddress,
 aps,pre]{revtex4-2}
\usepackage{amsmath}
\usepackage{amssymb}
\usepackage[dvipdfmx]{graphicx}
\usepackage{bm}
\usepackage{color} 
\begin{document}

\title{Deformation dynamics of an oil droplet into a crescent shape during intermittent motion}

\author{Sayaka Otani}\affiliation{Department of Physics, Chiba University, Yayoi-cho 1-33, Inage-ku, Chiba 263-8522, Japan}
\author{Hiroaki Ito}\affiliation{Department of Physics, Chiba University, Yayoi-cho 1-33, Inage-ku, Chiba 263-8522, Japan}
\author{Tomonori Nomoto}\affiliation{Department of Applied Chemistry and Biotechnology, Chiba University, Yayoi-cho 1-33, Inage-ku, Chiba 263-8522, Japan}
\author{Masanori Fujinami}\affiliation{Department of Applied Chemistry and Biotechnology, Chiba University, Yayoi-cho 1-33, Inage-ku, Chiba 263-8522, Japan}
\author{Jerzy G\'{o}recki}\affiliation{Institute of Physical Chemistry, Polish Academy of Sciences, Kasprzaka 44/52, Warsaw 01-224, Poland}
\author{Hiroyuki Kitahata}\email{kitahata@chiba-u.jp}\affiliation{Department of Physics, Chiba University, Yayoi-cho 1-33, Inage-ku, Chiba 263-8522, Japan}

\date{\today}

\begin{abstract}
    A paraffin droplet containing camphor and oil red O (dye) floating on the water surface shows spontaneous motion and deformation generated by the surface tension gradient around the droplet. We focused on the intermittent motion with a pronounced deformation into a crescent shape observed at specific concentrations of camphor and oil red O. We quantitatively analyzed the time changes in the droplet deformation and investigated the role of the oil red O by measuring the time-dependent paraffin/water interfacial tension with the pendant drop method. The observed effect can be explained by the active role of the oil red O molecules at the paraffin/water interface. The interfacial tension decreases gradually after the interface formation, allowing for the dynamic deformation of the droplet. The combination of the decrease in interfacial tension and the reduction in driving force related to camphor outflow generates intermittent motion with dynamic deformation into a crescent shape.
\end{abstract}
\maketitle

\section{\label{sec:level1}Introduction}

Self-propelled motion can be observed in many nonequilibrium systems, and it is regarded as one of the attributes of life\cite{Luisi}. There are also nonliving systems in which the free energy is converted into mechanical motion. They are much less complex than living organisms, and the mechanism of their self-propulsion can be explained with mathematical models through physics and chemistry. Consequently, studies on nonliving self-propelled motion are crucial for a deep understanding of the phenomenon of self-motion on the grounds of nonequilibrium science.

A camphor piece placed on the water surface is one of the well-known examples of self-propelled objects\cite{10.1039/9781788013499,C5CP00541H}. The self-propulsion of a camphor disk has been studied for more than a century\cite{doi:10.1098/rspl.1860.0124,doi:10.1098/rspl.1889.0099}. After a disk is placed on the water surface, a layer of camphor molecules is formed, reducing the surface tension. Fluctuations in the camphor surface concentration can appear around the disk, leading to the surface tension gradient that drives the disk toward the lower concentration area. Once the disk starts moving, its motion is maintained since the surface tension in front of a disk is higher\cite{KARASAWA2018184,doi:10.1246/cl.140201,doi:10.1021/acs.jpcb.3c00466}. Thus, a circular camphor disk moves by spontaneous symmetry breaking\cite{NAGAYAMA2004151,doi:10.1021/jp004505n}. In typical laboratory conditions, camphor molecules evaporate quickly; hence, the water surface tension recovers in a short time. Therefore, camphor disk motion can be observed for a long time exceeding one hour.  While fluctuations govern the direction of motion of a circular camphor disk, the motion of an asymmetric disk is determined by its shape\cite{doi:10.1021/la970196p,PhysRevE.99.062605,MOROHASHI2019104,D0SM01393E,doi:10.7566/JPSJ.89.094001,10.3389/fphy.2022.858791,PhysRevE.87.010901, iida2014theoretical}. For example, it has been demonstrated both theoretically and experimentally that an elliptical object tends to move in the direction of its minor axis\cite{PhysRevE.87.010901, iida2014theoretical}. There are also several other systems in which the particle shape determines the motion\cite{doi:10.1021/la801329g, AKELLA20181176,doi:10.1021/jp405364c, molecules26113116}. 

For the self-propelled motion of a solid object, its shape determines the motion. In the case of a soft, deformable, self-propelled object such as an oil droplet containing camphor, the motion is also initiated and supported by asymmetric concentration of surface active molecules around. However, for such an object, the feedback between motion and shape can be anticipated because the motion can modify the object shape, and the shape affects its motion. The interactions between motion and shape of self-propelled droplets were studied in a number of papers\cite{doi:10.1021/jp9616876,PhysRevLett.94.068301,PhysRevE.71.065301,10.1143/PTPS.161.286,BONIFACE2022990,https://doi.org/10.1002/anie.201104261,banno2016deformable,PhysRevE.76.055202,D3SM00228D,doi:10.1073/pnas.2121147119,ebata2015swimming,PhysRevLett.93.027802,doi:10.1021/la800105h}. The mechanism of the coupling between motion and deformation is an essential problem in self-propulsion in non-equilibrium systems. There have been several theoretical reports in which generic models for the coupling between motion and deformation by considering the symmetric properties were discussed\cite{PhysRevLett.102.154101, Tarama_2016,doi:10.7566/JPSJ.86.072001}. These generic models are also applicable to the biological cell motion with deformation\cite{OhtaSano,EbataKidoaki}. However, these studies were concerned with small magnitudes of deformation from a circle. As far as the authors know, there has been no universal theoretical description that can be applied to large deformation.

Experimental results on self-motion of a paraffin droplet with camphor floating on the water surface have been reported. Such a droplet shows complex behavior that can be controlled by the concentration of the oil red O in paraffin\cite{10.1162/isal_a_00106,Loeffler_Thesis}. For moderate oil red O concentrations, a droplet moves spontaneously with shape deformation that cannot be described as a small perturbation from a circular shape. A quantitative analysis of the phenomenon should help for better understanding of the self-propulsion in systems showing significant shape deformations.

In the presented study, we investigated the mechanism of deformation of a paraffin droplet containing camphor and oil red O, in which intermittent motion with dynamic deformation can be observed depending on the concentrations of camphor and oil red O and the elapsed time from putting the droplet on the water surface. We mainly focused on the change between the crescent and circular shapes observed during the intermittent motion. We introduced the quantities that allow us to measure the degree of deformation and evaluated them by image processing of the experimental results. Then, we discussed the relationship between these quantities and the droplet speed. We also measured the interfacial tension between water and paraffin containing oil red O and explained the deformation mechanism towards a crescent shape coupled with intermittent motion.

\section{\label{sec:level2}Experimental Setup}
\subsection{Spontaneous motion of a droplet}
(+)-Camphor and oil red O (dye) were purchased from FUJIFILM Wako Pure Chemical Corporation (Japan), and paraffin was purchased from Sigma-Aldrich (USA). The chemicals were used without further purification. Ion-free water was obtained using Elix UV3 (Merck, Germany). We prepared a paraffin solutions of camphor and oil red O at different concentrations by stirring using a vortex mixer and sonication.

We observed the time evolution of $20~\mu\mathrm{L}$ droplets located on the water surface in a Petri dish with an inner diameter of $194.5~\mathrm{mm}$. The time evolution of the system was recorded from above for an hour at $50~\mathrm{fps}$ with the CMOS camera (STC-MBS43U3V, OMRON SENTECH Co., LTD., Japan) equipped with the objective lens (3Z4S-LE SV-0814H, OMRON Co., LTD., Japan). The Petri dish was illuminated by a light panel placed below. 

All experiments were performed at room temperature ($22 \pm 2{}^\circ \mathrm{C}$). The recorded videos were analyzed using an image-processing software ImageJ\cite{ImageJ}.

\subsection{Interfacial tension between oil and water}
The interfacial tension between pure water and oil was measured using the pendant drop method. Paraffin containing oil red O was ejected through a hook-shaped needle (4983 PD 22G, Kyowa Interface Science Co., Japan) into the aqueous phase. The droplet image was recorded from the side with the above-mentioned camera equipped with the objective lens (TEC-M55, CBC Co., Ltd., Japan). It was illuminated by the LED lamp (AS3000, AS ONE Co., Japan) located at the opposite side to the camera. The interfacial tension was calculated by fitting the shape of an oil droplet to a theoretical curve. The details are described in Appendix~\ref{app:pendant_drop}.

\section{\label{sec:level3}Results}
\subsection{Spontaneous motion of a droplet}
\begin{figure}[tb]
    \includegraphics{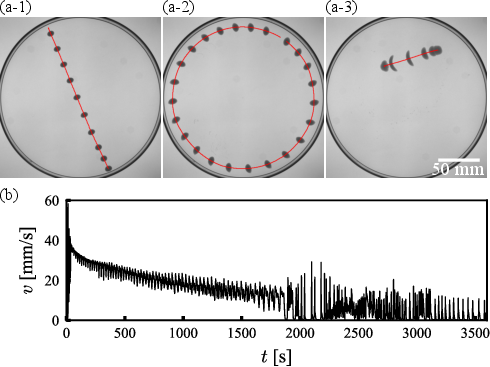}
    \caption{\label{fig:trajectory} 
    Typical long-time behavior of a paraffin droplet containing camphor (concentration $0.06~\mathrm{M}$) and oil red O (concentration $1.0~\mathrm{g/L}$) moving on a water surface. The time $t = 0$ corresponds to the moment when the droplet was placed on the water surface. (a) Snapshots of the droplet motion at around $t \sim 0~\mathrm{s}$ (a-1), $t \sim 600~\mathrm{s}$ (a-2), and $t \sim 2100~\mathrm{s}$ (a-3). The superposed images of the droplet are shown with the time difference of $0.5~\mathrm{s}$ (a-1) and $1.0~\mathrm{s}$ (a-2, a-3). The red lines indicate the trajectories of the droplet center of mass (COM) position. (b) Droplet speed $v$ as a function of time $t$.}
\end{figure}

The motion of a $20~\mu\mathrm{L}$ droplet of paraffin with dissolved camphor (concentration $0.06~\mathrm{M}$) and oil red O (concentration $1.0~\mathrm{g/L}$) clearly demonstrated the transient character of the phenomenon. Three different types of motion could be easily distinguished for different times $t$ elapsed from the moment when the droplet was placed on the water surface. At the beginning of the experiment, the droplet showed self-propelled motion along a straight line perturbed by reflections with the dish wall, as illustrated in Fig.~\ref{fig:trajectory}(a-1). Approximately at $t=30~\mathrm{s}$, the droplet began to rotate along the wall (Fig.~\ref{fig:trajectory}(a-2)). Such type of motion was observed for over $1500~\mathrm{s}$. The droplet speed began to oscillate with an increasing amplitude and a decreasing average as illustrated in Fig.~\ref{fig:trajectory}(b). Then, at $t \sim 2000~\mathrm{s}$, the droplet motion switched to intermittent one (Fig.~\ref{fig:trajectory}(a-3)). At this stage, short intervals of time when a droplet moved (less than $10~\mathrm{s}$) were separated by periods when it was standing. The droplet had a circular shape when it did not move, and it became elongated and exhibited deformation into a crescent shape during intermittent motion. We observed a monotonic decrease in the peak speed as a function of time. It can be anticipated that the changes in the character of motion are related to the rate of camphor release, which decreases the water surface tension around it. However, as we discuss below, the slow modification of interfacial tension of the oil-water interface resulting from the presence of dye molecules near the surface is another factor leading to intermittent motion.

\begin{figure}[tb]
    \includegraphics{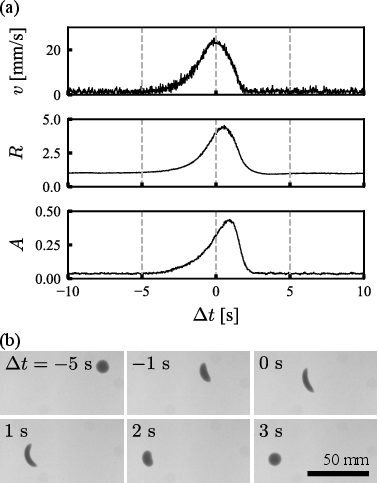}
    \caption{\label{fig:A_r_v}
    Dynamics of a single phase of the intermittent motion of a $20~\mu\mathrm{L}$ paraffin droplet with camphor and oil red O. The concentrations of camphor and oil red O were $0.06~\mathrm{M}$ and $1.0~\mathrm{g/L}$, respectively. (a) The speed $v$ of the droplet COM, the aspect ratio $R$, and the asymmetric parameter $A$ as functions of $\Delta t$. The time $\Delta t = 0$ corresponds to when $v$ has the maximum value. (b) A sequence of frames showing droplet location and deformation during the phase of intermittent motion analyzed in (a).}
\end{figure}

Hereafter, we focus on the intermittent motion shown in Fig.~\ref{fig:trajectory}(a-3). We defined a single phase of intermittent motion based on the droplet speed as a function of time (cf. Fig.~\ref{fig:A_r_v}(b)). If the droplet stopped ($v\leq 2~\mathrm{mm/s}$) for more than $4~\mathrm{s}$, then we assumed that the droplet was at the rest state. If the droplet at the rest state accelerated and its speed exceeded $10~\mathrm{mm/s}$, then we considered this moment as the start of the intermittent phase. In the case that the time duration between the start of motion and the return to the rest state was less than $10~\mathrm{s}$, we defined that the droplet had performed an intermittent motion, and we analyzed the time series during each intermittent motion in detail. To minimize the boundary effect we did not analyze the intermittent motion if the droplet got closer than $20~\mathrm{mm}$ to the wall of the Petri dish. In case that a droplet exhibited fission, we excluded the data after the fission.

\begin{figure}[tb]
    \includegraphics{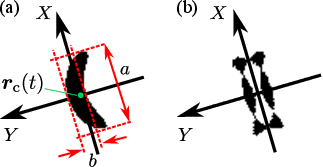}
    \caption{
    \label{fig:parameters} 
    Illustration of parameters describing the droplet deformation. (a) Binarized image of a droplet. Aspect ratio is defined as $R = a/b$. (b) Image obtained by applying XOR operation to the droplet image and the image reflected to the $X$-axis. Asymmetric parameter $A$ is defined as the area of the image normalized by twice the area of the original image.}
\end{figure}

In the following analysis, we focused on the correlations between the droplet speed and dynamic deformation into a crescent shape observed during the intermittent motion (Fig.~\ref{fig:trajectory}(a-3)). The observed droplet elongation was always perpendicular to the direction of motion. For quantitative study on the deformation dynamics, we introduced the local coordinates associated with the direction of motion. Each frame of the observed time evolution was processed so that pixels representing the droplet were black (the logical value $1$) and all the others were white (the logical value $0$). For each frame, we calculated the droplet center $\bm{r}_{\mathrm{c}}(t)$ as the center of mass (COM) of the black pixels. The set of droplet centers defined the droplet trajectory. The origin of the local coordinate system was set at $\bm{r}_{\mathrm{c}}(t)$. The $Y$-axis was defined as parallel to a fitting line to the droplet trajectory with a least-square method and directed as the net displacement vector. The $X$-axis was set perpendicular to the $Y$-axis. Using these coordinates, we defined the time-dependent aspect ratio $R = a/b$, where $b$ is the width of the droplet measured on the $Y$-axis and $a$ is the length of the droplet projected on the $X$-axis as illustrated in Fig.~\ref{fig:parameters}(a). The second quantity describing time-dependent droplet shape is the front-back asymmetric parameter $A = \Delta \Sigma / (2\Sigma)$, where $\Sigma$ is the area (the number of black pixels) of the droplet image and $\Delta \Sigma$ is the area of the figure obtained by applying XOR operation to the original droplet image and its image reflected with respect to the $X$-axis as shown in Fig.~\ref{fig:parameters}(b). It should be noted that $0 \leq A \leq 1$ and $A=0$ for a droplet with mirror symmetry on the $X$-axis.

Figure \ref{fig:A_r_v}(a) shows the speed $v(t)$ of the droplet center $(\bm{r}_{\mathrm{c}}(t))$ and the deformations $R(t)$ and $A(t)$ as functions of time during a single phase of intermittent motion illustrated in Fig.~\ref{fig:A_r_v}(b). The time $\Delta t = 0$ corresponds to the moment when $v(t)$ has the maximum value. As shown in Fig.~\ref{fig:A_r_v}(a), the deformations began almost simultaneously when motion started. The speed $v(t)$ reached its maximum before the maxima of $R(t)$ and $A(t)$, which means that deformations still developed after the droplet started to slow down. The comparison between $v(t)$, $R(t)$ and $A(t)$ shows that the maximum acceleration appeared when the values of both the parameters describing deformation were around $50\%$ of their maximum.

\begin{figure}[tb]
    \includegraphics{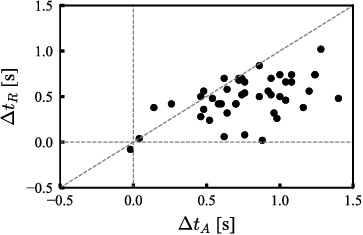}
    \caption{\label{fig:dt_hist}
    Scatter plot of $\Delta t_A$ and $\Delta t_R$, which are times for maximum values of $A(t)$ and $R(t)$ for the paraffin droplet with camphor and oil red O. The camphor concentration was $0.06~\mathrm{M}$ and oil red O concentration was $1.0~\mathrm{g/L}$.}
\end{figure}

The behavior illustrated in Fig. \ref{fig:A_r_v}(a)  was reproduced in many observations of intermittent motion, as shown in Fig.~\ref{fig:dt_hist}, in which the scatter plot of the time $\Delta t_R$ for the maximum of $R(t)$ and $\Delta t_A$ for the maximum of $A(t)$ is displayed. In almost all analyzed cases, both $\Delta t_R$ and $\Delta t_A$  were positive, and $\Delta t_A$ was slightly greater than $\Delta t_R$. 

\begin{figure}[tb]
    \includegraphics{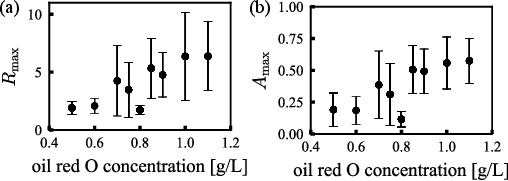}
    \caption{\label{fig:oro_conc_deformation_param} 
    (a) Maximum values of $R_{\mathrm{max}}$ and (b) $A_{\mathrm{max}}$, for $R(t)$ and $A(t)$ as functions of oil red O concentration. The camphor concentration was $0.06~\mathrm{M}$, and experiments at each oil red O concentration were repeated more than $5$~times. The error bars indicate the standard deviation of the results.}
\end{figure}

Next, we investigated the dependence of the droplet deformation on the oil red O concentration. We performed experiments at oil red O concentrations in the range of $0.5$--$1.1~\mathrm{g/L}$. The maximum values of $R(t)$ and $A(t)$ ($R_{\mathrm{max}}$ and $A_{\mathrm{max}}$) increased for oil red O concentrations larger than $0.8~\mathrm{g/L}$ as shown in Fig.~\ref{fig:oro_conc_deformation_param}.

\subsection{Interfacial tension between oil and water\label{subsec:experiment2}}

\begin{figure}[tb]
    \includegraphics{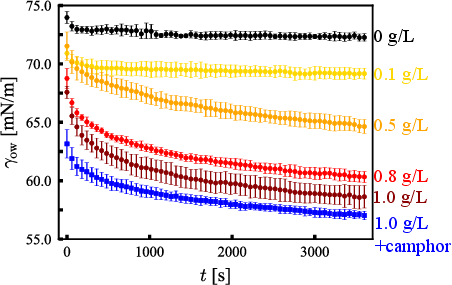}
    \caption{\label{fig:ift1}
    Time evolution of the interfacial tension $\gamma_{\mathrm{ow}}(t)$ between a paraffin droplet and pure water measured by the pendant drop method. The time $t=0$ corresponds to when a droplet was formed.  Marker colors code the oil red O concentrations: $0~\mathrm{g/L}$ black, $0.1~\mathrm{g/L}$ yellow, $0.5~\mathrm{g/L}$ orange, $0.8~\mathrm{g/L}$ red, $1.0~\mathrm{g/L}$ dark red. For each concentration, the average values of four experiments are plotted. The error bars indicate the standard deviation of the results. $\gamma_{\mathrm{ow}}(t)$ between pure water and paraffin droplet that contained both oil red O ($1.0~\mathrm{g/L}$) and camphor ($0.06~\mathrm{M}$) is marked by blue squares.}
\end{figure}

\begin{figure}[tb]
    \includegraphics{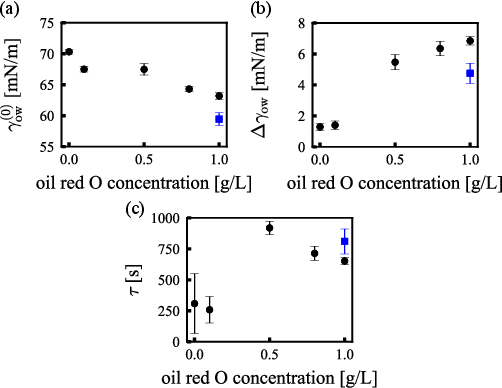}
    \caption{\label{fig:ift_conc} 
    Parameters fitting $\gamma_{\mathrm{ow}}(t)$ according to Eq.~\eqref{eq:ift_fitting} for different concentrations of oil red O (cf. Fig.~\ref{fig:ift1}). (a) Interfacial tension on the initial stage $\gamma_{\mathrm{ow}}^{(0)}$. (b) Difference in interfacial tension $\Delta\gamma_{\mathrm{ow}}$ between the initial and final stages. (c) Characteristic time $\tau$ for the decrease in interfacial tension.
    The data for a paraffin droplet that contained both oil red O ($1.0~\mathrm{g/L}$) and camphor ($0.06~\mathrm{M}$) are marked with blue squares.
    The average values for four measurements are shown. The error bars indicate their standard deviations.}
\end{figure}

To understand the role of oil red O during the droplet evolution, we measured the interfacial tension $\gamma_{\mathrm{ow}}(t)$ between a paraffin droplet containing oil red O and pure water using the pendant drop method\cite{doi:10.1098/rspa.1948.0063} as a function of time. The method is described in Appendix~\ref{app:pendant_drop}. The measurements were performed for various oil red O concentrations. For the oil red O concentration $\geq 0.5~\mathrm{g/L}$, the interfacial tension gradually decreased in time after the interface was formed as shown in Fig.~\ref{fig:ift1}. Assuming that the diffusional transport of oil red O molecules and the linear absorption process of them at the droplet interface are responsible for the decrease in the interfacial tension, the time evolution of the interfacial tension should be described as an exponential relaxation. Thus, for quantitative evaluation, we fitted time-dependent interfacial tension $\gamma_{\mathrm{ow}}(t)$ to an exponential function 
\begin{align}
    \gamma_{\mathrm{ow}}(t)=\gamma_{\mathrm{ow}}^{(0)}-\Delta\gamma_{\mathrm{ow}}(1-e^{-t/\tau}),\label{eq:ift_fitting}
\end{align}
where $\gamma_{\mathrm{ow}}^{(0)}$ is the interfacial tension at the initial stage, $\Delta\gamma_{\mathrm{ow}} = (\gamma_{\mathrm{ow}}^{(0)}-\gamma_{\mathrm{ow}}(\infty)) $  is the difference in interfacial tension between the initial and final stages , and $\tau$ is the characteristic time for the decrease in interfacial tension. This functional form comes from the model describing the rate of relaxation of interfacial tension towards its stable value as proportional to $(\gamma_{\mathrm{ow}}(\infty) - \gamma_{\mathrm{ow}}(t))$. Such behavior can be expected when the dye is absorbed on the droplet surface. We plotted $\gamma_{\mathrm{ow}}^{(0)}$, $\Delta\gamma_{\mathrm{ow}}$ and $\tau$ against oil red O concentration in Fig.~\ref{fig:ift_conc}. $\gamma_{\mathrm{ow}}^{(0)}$ and $\Delta\gamma_{\mathrm{ow}}$ were decreasing and increasing functions of oil red O concentration, respectively. In the case when the decrease in interfacial tension was clearly observed (oil red O concentrations: $0.5$, $0.8$, and $1.0~\mathrm{g/L}$), $\tau$ was approximately equal to $800~\mathrm{s}$. In addition, we performed the same measurement for the oil droplet that contained both oil red O ($1.0~\mathrm{g/L}$) and camphor ($0.06~\mathrm{M}$). The decrease in interfacial tension was also observed, and $\gamma_{\mathrm{ow}}^{(0)}$ was lower compared to paraffin droplet with oil red O ($1.0~\mathrm{g/L}$) but without camphor.

\section{\label{sec:level4}Discussion}

\subsection{Analysis of free energy related to droplet deformation\label{subsec:discuttion1}}

To better understand the droplet deformation phenomenon, we evaluated two related dimensionless numbers. First, the capillary number $\mathrm{Ca}$ is the ratio of the viscous force and interfacial tension, $\mathrm{Ca} = \mu v/\gamma_{\mathrm{ow}}$, where $\mu$ is the viscosity of the liquid, $v$ is the characteristic velocity, and $\gamma_{\mathrm{ow}}$ is the characteristic value for oil/water interfacial tension. By substituting $\mu \simeq 2 \times 10^{-1}~\mathrm{Pa\,s}$ (paraffin), $v \simeq 10^{-2}~ \mathrm{m/s}$, and $\gamma_{\mathrm{ow}} \simeq 60 \times 10^{-3}~\mathrm{N/m}$, we obtain $\mathrm{Ca} \simeq 3 \times 10^{-2} \ll 1$. This means that during an intermittent motion, the relative effect of viscous drag forces is much smaller than that of surface tension.

Next, the Weber number $\mathrm{We}$ is the relative importance of the fluid inertia compared to its surface tension. $\mathrm{We} = \rho r v^2/\gamma_{\mathrm{ow}}$, where $\rho$ is the mass density of the liquid and $r$ is the characteristic length. For $\rho \simeq 10^3~\mathrm{kg/m^3}$, $r \simeq 10^{-2}~\mathrm{m}$ and other parameter values listed above, we obtained $\mathrm{We} \simeq 2\times 10^{-5} \ll 1$. This means that the inertial effects are negligible during an intermittent motion. Considering both the dimensionless numbers, we concluded that the surface tension difference around the droplet plays the most important role in droplet deformation.

\begin{figure}[tb]
    \includegraphics{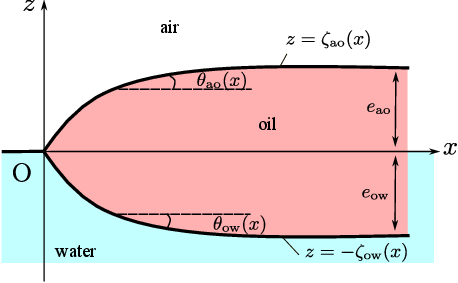}
    \caption{\label{fig:tripleline} 
Schematic illustration of an oil droplet floating on the water surface. The triple line matches $x=z=0$. The functions $z = \zeta_{\mathrm{ao}}(x)$ and $z = -\zeta_{\mathrm{ow}}(x)$ indicate the interface between air and oil and that between oil and water, respectively. $\theta_{\mathrm{ao}}(x)$ and $\theta_{\mathrm{ow}}(x)$ determine the slopes of them, i.e., $\tan\theta_{\mathrm{ao}}(x)=\mathrm{d}\zeta_{\mathrm{ao}}/\mathrm{d}x$ and $\tan\theta_{\mathrm{ow}}(x)=\mathrm{d}\zeta_{\mathrm{ow}}/\mathrm{d}x$. $e_{\mathrm{ao}}$ and $e_{\mathrm{ow}}$ are the differences in height between the water level at infinity and the air/oil interface, and that between the water level at infinity and the oil/water interface, respectively.}
\end{figure}

In the following, we use the approximation of droplet free energy to derive the energy condition for a spontaneous deformation of droplet shape. If we neglect the contribution of the periphery, then the free energy of a floating droplet shown in Fig.~\ref{fig:tripleline} is estimated as:
\begin{align}
    F  = -S\Sigma + \frac{1}{2}\left(\Delta\rho_{\mathrm{ao}}e_{\mathrm{ao}}^2+\Delta\rho_{\mathrm{ow}}e_{\mathrm{ow}}^2\right)g\Sigma, \label{Eq_F}
\end{align}
Here, $S = \gamma_{\mathrm{aw}}-(\gamma_{\mathrm{ow}}+\gamma_{\mathrm{ao}})$ is the spreading coefficient, $\gamma_{\mathrm{aw}}$ and $\gamma_{\mathrm{ao}}$ are the interfacial tensions between air and water and between air and oil, respectively, $g$ is the gravitational acceleration, $\Sigma$ is the area of the droplet, and $\Delta\rho_{\mathrm{ao}}$ and $\Delta\rho_{\mathrm{ow}}$ are the density differences between air and oil and that between oil and water, respectively, as shown in Fig.~\ref{fig:tripleline}. In Eq.~\eqref{Eq_F}, $e_{\mathrm{ao}}$ and $e_{\mathrm{ow}}$ are the differences in height between the water level at infinity and the air/oil interface, and that between the water level at infinity and the oil/water interface, respectively. For the following analyses, we neglected the second term in Eq.~\eqref{Eq_F} as it is much smaller than the first term since the thickness of a droplet is of the order of $100~\mu\mathrm{m}$, thus much smaller than the droplet size.

The second contribution to droplet energy comes from the curvature of its outer edge. The excess energy of the triple line per unit length equals:
\begin{align}
    \mathcal{T}
    =&\frac{4}{3}\gamma_{\mathrm{ao}}\kappa_{\mathrm{ao}}^{-1}\left[1-\cos^2\left(\frac{\theta_{\mathrm{ao}}(x=0)}{2}\right)\right] \label{eq:line_tension}\\
    &+ \frac{4}{3}\gamma_{\mathrm{ow}}\kappa_{\mathrm{ow}}^{-1}\left[1-\cos^2\left(\frac{\theta_{\mathrm{ow}}(x=0)}{2}\right)\right],\nonumber
\end{align}
where $\kappa_{\mathrm{ao}}^{-1}$ and $\kappa_{\mathrm{ow}}^{-1}$ are capillary lengths of air/oil and oil/water interfaces, respectively. $\theta_{\mathrm{ao}}(x=0)$ is the angle between the air/oil interface and the $x$-axis at the triple line, and  $\theta_{\mathrm{ow}}(x=0)$ is that between the oil/water interface and the $x$-axis. We calculated the excess energy of the triple line using the relationships linking the angles $\theta_{\mathrm{ao}}(x=0)$,  $\theta_{\mathrm{ow}}(x=0)$ with the interfacial tensions derived in Appendix~\ref{appsub:line_tension}.

The change in free energy related to droplet deformation can be approximated as:
\begin{align}
    \Delta E
    = -S\Delta \Sigma + \mathcal{T}\Delta l,
\end{align}
where the first term is the energy for an increase in droplet area by $\Delta \Sigma$, and the second describes the energy for an increase in droplet periphery $\Delta l$. If the surface tension difference between the front and back of the droplet is $\delta\gamma_{\mathrm{aw}}$ and the droplet moved on a water surface by an area of $\sigma$, then the work exerted on the droplet is estimated as
\begin{align}
    W = \delta\gamma_{\mathrm{aw}}\sigma.
\end{align}

\begin{figure}[tb]
    \includegraphics{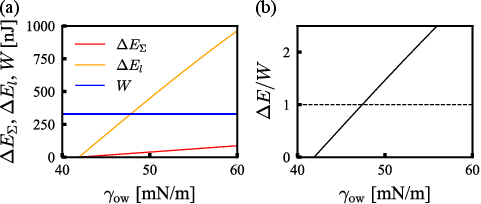}
    \caption{\label{fig:ift_energy}
   Major forms of energy related to droplet deformation as functions of the interfacial tension between oil and water. (a) Plots of $\Delta E_{\Sigma}$ (red), $\Delta E_l$ (orange) and $W$ (blue), respectively. (b) Solid line shows deformation energy normalized by the work exerted on the droplet, $\Delta E/W$ against $\gamma_{\mathrm{ow}}$.}
\end{figure}

To support self-motion with deformation, the energy gain due to the surface tension gradient should exceed the free energy increase due to the droplet deformation depending on the interfacial tension. If the excess energy of the triple line dominates, the system relaxes towards a circular shape as, for example, demonstrated in the case of soap films\cite{PRL_Lenavetier}. In our system, the quasi-elastic laser scattering (QELS) measurements shown in Appendix~\ref{app:QELS} suggested that the value of $\delta \gamma_{\mathrm{aw}}$ does not exceed $0.3~\mathrm{mN/m}$, though we were not able to measure this value directly. For the following calculations, we used $\delta\gamma_{\mathrm{aw}}=0.25~\mathrm{mN/m}$. During the intermittent motion shown in Fig.~\ref{fig:A_r_v}, we observed $\Delta \Sigma = 3.3~\mathrm{mm}^2$, $\Delta l = 17.2~\mathrm{mm}$ and $\sigma = 1288.5~\mathrm{mm}^2$. For these parameters and values of interfacial tensions measured by the Wilhelmy method: $\gamma_{\mathrm{ao}}=31.6~\mathrm{mN/m}$ and $\gamma_{\mathrm{aw}}=73.5~\mathrm{mN/m}$, the values of  $\Delta E_{\Sigma} = S\Delta \Sigma$, $\Delta E_l = \mathcal{T} \Delta l$ and $W$ are plotted as functions of $\gamma_{\mathrm{ow}}$ in Fig.~\ref{fig:ift_energy}(a). For these realistic values of parameters ($\gamma_{\mathrm{ow}}\sim 50~\mathrm{mN/m}$), $\Delta E_{\Sigma}+\Delta E_l$, and $W$ are of the same order, which implies that the deformation of a droplet can occur. $\Delta E_{\Sigma}(\gamma_{\mathrm{ow}})$ and $\Delta E_l(\gamma_{\mathrm{ow}})$ are increasing functions of interfacial tension $\gamma_{\mathrm{ow}}$, while $W$ does not depend on it. The increase in $\Delta E_l(\gamma_{\mathrm{ow}})$ is faster than that in $\Delta E_{\Sigma}(\gamma_{\mathrm{ow}})$, which means that the line energy is the most important for droplet deformation. Figure~\ref{fig:ift_energy}(b) shows the deformation energy normalized by the work exerted on the droplet. For the selected parameter values, the spontaneous droplet deformation can occur when $\Delta E/W <1 $, which happens if $\gamma_{\mathrm{ow}} \lesssim 47~\mathrm{mN/m}.$ Having in mind that $\gamma_{\mathrm{ow}}$ is a decreasing function of oil red O concentration (cf. Fig.~\ref{fig:ift1}), we can see why droplet deformation occurs for droplets with a higher concentration of oil red O. The estimated threshold for deformation is a bit lower than expected from the results discussed in the previous chapter. There could be many reasons for this. First, the estimation is based on a simple model. Second, the local inhomogeneities in oil red O concentration, especially those near the droplet boundary, can trigger local shape changes. And finally, there may be possible errors in the measurements of interfacial tension. For example, the threshold should increase for the lower value of $\gamma_{\mathrm{ao}}$.

\subsection{\label{subsec:dynamics}Dynamics of droplet deformation induced by camphor release}

\begin{figure}[tb]
    \includegraphics{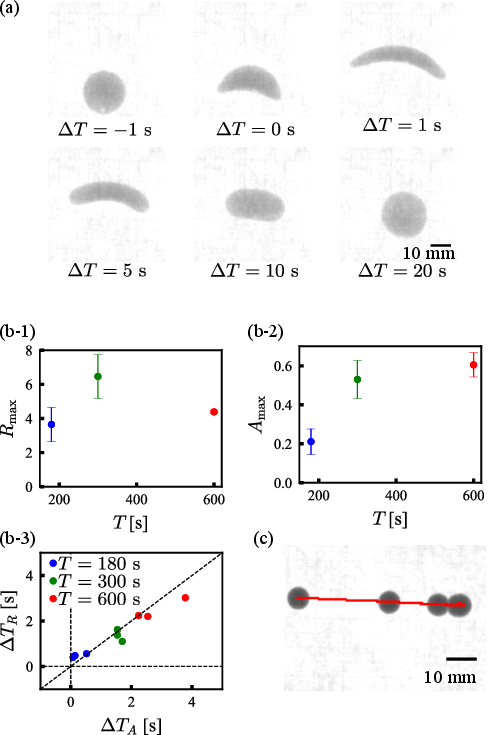}
    \caption{\label{fig:camphordisk}
    Deformation of a paraffin droplet containing oil red O (concentration $1.0~\mathrm{g/L}$) observed when the water surface was touched with a camphor disk at time $T$ after the droplet had been placed on the water surface. The time $\Delta T=0$ corresponds to when $v$ has the maximum value. (a) Snapshots of a droplet for $T=300~\mathrm{s}$. (b) The values of parameters characterizing droplet deformation. Panels (b-1) and (b-2) show the average value for $R_{\mathrm{max}}$ and $A_{\mathrm{max}}$ with the standard deviation as error bars for 3 experiments for each $T$, and (b-3) is a scatter plot of $\Delta T_A$ and $\Delta T_R$. (c) Superposed images of a droplet for $T=30~\mathrm{s}$ with a time difference of $2.0~\mathrm{s}$. The red line indicates the trajectories of the droplet COM.}
\end{figure}

The discussion of energetic aspects of droplet deformation indicates that a deformation occurs  easier for a long time $t$ after the droplet is placed on the water surface. To verify this observation, we investigated deformations of paraffin droplets with oil red O that do not contain camphor. Of course, if such droplets are placed on the water surface they do not move and do not change their shapes. The surface tension gradient necessary for the motion was induced by touching the water surface close to the droplet with a camphor piece at time $T$ after the droplet had been placed on water. The experiment with a $20~\mu\mathrm{L}$ paraffin droplet containing oil red O (concentration $1.0~\mathrm{g/L}$) is illustrated in Fig.~\ref{fig:camphordisk}. As a camphor source, we used a $3~\mathrm{mm}$-diameter camphor disk located around $5~\mathrm{mm}$ away from the droplet boundary. After the camphor disk touched the surface, the droplet moved away from the source with deformation into a crescent shape, as shown in Fig.~\ref{fig:camphordisk}(a). We performed experiments for $T = 30, 180, 300, 600~\mathrm{s}$ and plotted the characteristic parameters describing the droplet shape dynamics in Fig.~\ref{fig:camphordisk}(b). The time $\Delta T=0$ corresponds to when the droplet COM speed $v$ has the maximum value. The maximum magnitude of deformation $A_{\mathrm{max}}$ increased with $T$ as shown in Fig.~\ref{fig:camphordisk}(b-2). The shape of the droplet was the most curved when $T=600~\mathrm{s}$, which can explain why $R_{\mathrm{max}}(T = 600~\mathrm{s})$ was smaller than that when $T=300~\mathrm{s}$. Figure~\ref{fig:camphordisk}(b-3) shows a scatter plot of the times $\Delta T_R$ and $\Delta T_A$, at which the maximum values of $R$ and $A$ ($R_{\mathrm{max}}$ and $A_{\mathrm{max}}$) were observed. Both $\Delta T_R$ and $\Delta T_A$ were positive and had almost the same value. This result is similar to those for the spontaneous deformation of a droplet with camphor shown in Fig.~\ref{fig:dt_hist}, though the relationship between $\Delta T_R$ and $\Delta T_A$ was slightly different from that between $\Delta t_R$ and $\Delta t_A$. However, the times $\Delta T_R$ and $\Delta T_A$ were over twice as large as $\Delta t_R$ and $\Delta t_A$. The difference can be attributed to the larger surface concentration gradient in the case of a camphor-free droplet than the one appearing for the paraffin droplet with camphor. When $T=30$~s, the droplet hardly deformed, as shown in Fig.~\ref{fig:camphordisk}(c), and the characteristic parameters could not be appropriately obtained. These results indicate that for longer $T$ of the order of $10$~minutes, the modification of the droplet interface becomes more efficient, leading to more pronounced droplet deformation.

\subsection{\label{subsec:flows}Dynamics of  flows around the droplet}

In this subsection, we focus our attention on time-dependent velocity and deformation during the intermittent droplet motion. In particular, we focus our attention on the role of the oil red O because in the previous study, a large deformation of a droplet shape without fission has not been observed without oil red O\cite{10.1162/isal_a_00106}.

The results shown in Figs.~\ref{fig:A_r_v}, \ref{fig:dt_hist} and \ref{fig:camphordisk} indicate complex interactions between camphor and oil red O. For a droplet forced to move by an external source, $\Delta T_A$ and $\Delta T_R$ seem perfectly correlated. In this case, the time difference between the peak of velocity and that of the deformation can be as long as $4~\mathrm{s}$ and increases with time $T$. For a droplet containing camphor, $\Delta t_A$ and $\Delta t_R$ are much shorter ($< 1.5~\mathrm{s}$). Moreover, $\Delta t_R < \Delta t_A$, but the values are weakly correlated. The different relationship between the peaks can be attributed to the presence of surface active molecules inside a droplet.

To learn more about the intermittent motion mechanism and understand the relationship between time-dependent droplet velocity and shape deformation (cf. Figs.~\ref{fig:A_r_v} and \ref{fig:dt_hist}), we observed surface flows around the droplet exhibiting an intermittent motion. The flows were traced by small plastic particles (HP20SS, DIAION SEPABEADS, Mitsubishi Chemical, Japan), whose diameters were in the range of $60$--$150~\mu\mathrm{m}$ and the density was $1.01~\mathrm{kg/L}$. The particles were floating at the water surface due to the hydrophobic properties of their surfaces.

\begin{figure}[tb]
    \includegraphics{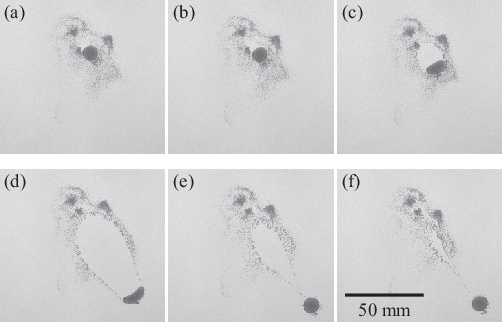}
    \caption{\label{fig:flow} Snapshots of floating small particles during an intermittent motion of a paraffin droplet (volume $20~\mu\mathrm{L}$) containing camphor (concentration $0.06~\mathrm{M}$) and oil red O (concentration $0.8~\mathrm{g/L}$) illustrating the surface flows around the droplet via locations of tracer particles. The snapshot (a) was made $1500~\mathrm{s}$ after placing the droplet on the water surface. The time difference between subsequent snapshots is $2~\mathrm{s}$.}
\end{figure}

The snapshots illustrating the evolution of droplets and flows are shown in Fig.~\ref{fig:flow}. Before the droplet started the intermittent motion, the particles were still on the water surface, which means that no strong surface flow was induced around the droplet (see Fig.~\ref{fig:flow}(a)). The evolution towards intermittent motion should be ruled by fluctuations in camphor release. A small region without tracer particles can be seen up-left of the droplet. The tracer particle-free area grows in time as seen in Figs.~\ref{fig:flow}(b)--(c) and generates a difference in surface tension between the upper and the lower boundaries of the droplet. The resulting force is large enough to drive the droplet towards the region with low camphor concentration. After more camphor molecules are released, the camphor-rich area (the one without tracer particles) grows large and contacts with almost a half of the droplet periphery. At this stage, the droplet acceleration reaches the maximum (Figs.~\ref{fig:flow}(c)--(d)). The droplet moves to the area with low camphor concentration and, as a consequence, the gradient of surface tension becomes low. In this phase of motion, the droplet slowed down and returned to the disk shape as shown in Figs.~\ref{fig:flow}(e)--(f).

In the experiments, the continuous motion changed to the intermittent motion circa half an hour after a droplet was placed on the water surface as shown in Fig.~\ref{fig:trajectory}. In the continuous motion, asymmetric concentration driving the droplet is supported by coupling between motion and high rate of camphor release. The transition to intermittent motion can be explained by combination of two factors. One is a decrease in camphor release rate due to declining camphor concentration in the droplet. However, the decrease alone does not fully explain the transition because paraffin droplets with small camphor concentration did not show intermittent motion. Our analysis shows that the modification of the water-oil interface by the presence of oil red O molecules is the factor allowing for moderation of camphor release as well as for changes in droplet shape. The release rate may be additionally affected by the flow around the droplet\cite{Kovalchuk}. For the clarification of the more detailed mechanism of this droplet deformation phenomenon, we need more precise measurement of the dynamics of camphor and oil red O molecules. We also need to construct a mathematical model that reproduces the time evolution of motion and deformation including hydrodynamics in future study.

\section{\label{sec:level5}Conclusions}
We experimentally observed that a paraffin droplet containing camphor and oil red O exhibits spontaneous motion with deformation. At considered concentrations of camphor and oil red O, the droplet started to perform intermittent motion approximately $2000~\mathrm{s}$ after placing it on the water surface. We defined the variables characterizing large deformation and evaluated the relationships between the velocity and deformation magnitude. We also measured the time change in oil/water interfacial tension and found that it gradually decreased in time due to the adsorption of oil red O at the oil/water interface. Simultaneously, the camphor concentration in the droplet should also decrease with time, resulting in a smaller driving force. The combination of decreased interfacial tension due to oil red O and reduced driving force due to reduced camphor concentration in the droplet causes intermittent motion with dynamic deformation into a crescent shape.

\begin{acknowledgments}
    We would like to thank Dr.~Richard~J.~G.~L\"{o}ffler for the fruitful discussion. This work was supported by PAN-JSPS program ``Complexity and order in systems of deformable self-propelled objects'' (No.~JPJSBP120234601). It was also financed by JSPS KAKENHI Grants Nos. JP21K13891 (H.I.), JP23K04687 (T.N.), JP20H02712, JP21H00996, and JP21H01004 (H.K.). This work was also supported by the JSPS Core-to-Core Program “Advanced core-to-core network for the physics of self-organizing active matter (JPJSCCA20230002, H.I. and H.K.) and Iketani Science and Technology Foundation(0351087-A, T.N.).
\end{acknowledgments}

\appendix

\section{\label{app:pendant_drop}Pendant drop method}
\subsection{Shape of a droplet}

\begin{figure}[tb]
    \centering
    \includegraphics{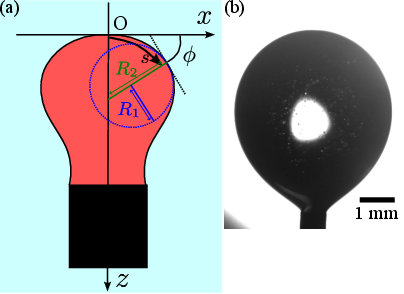}
    \caption{\label{fig:coordinate}Shape of the oil droplet on the tip of a needle. (a) Schematic illustration of the coordinate system used to describe a droplet. The red and cyan areas represent the oil and water phases, and the black marks the needle tip. The blue line marks the curvature radius $R_1$ in the $xz$-plane, and the green one marks the curvature radius $R_2$ in the plane perpendicular to the $xz$-plane. (b) The droplet image obtained in the experiment for the measurement with the pendant drop method.}
\end{figure}

Let us consider a droplet formed above the tip of a needle as shown in Fig.~\ref{fig:coordinate} since the density of the droplet is lower than that of the bulk. We set the Cartesian coordinates so that the $z$-axis corresponds to the center axis of the needle as shown in Fig.~\ref{fig:coordinate}(a). The top point of the droplet is set to be the origin, and the direction of $z$ is set downward. We assume the shape of the droplet is symmetric with respect to the $z$-axis. Due to the symmetry, we only consider the droplet profile on the $xz$-plane.

The balance between Laplace pressure and hydrostatic pressure yields
\begin{align}\label{eq:balance}
    \left(\frac{1}{R_1}+\frac{\sin\phi}{x}\right)\gamma = p_0 - \rho gz
\end{align}
at every point on the outline of the droplet except for the origin, where $R_1$ is a radius of curvature in the $xz$-plane, $\phi$ is an angle between the tangent line of the curve and the $x$-axis, $p_0$ is the hydrostatic pressure at the origin, and $\rho$ is a difference in density between the droplet and bulk. At the origin, the relation
\begin{align}\label{eq:origin}
\frac{2\gamma}{b} = p_0
\end{align}
is satisfied, where $b$ is the radius of curvature at the origin. Eliminating $p_0$ from Eqs. \eqref{eq:balance} and \eqref{eq:origin}, we get
\begin{align}
    \frac{1}{R_1} + \frac{\sin\phi}{x} = \frac{2}{b} - \frac{\rho g}{\gamma}z.
\end{align}
By introducing the length $s$ along the curve from the origin, we obtain
\begin{align}
    \frac{\mathrm{d}\phi}{\mathrm{d}s} = \frac{1}{R_1} = \frac{2}{b} - \beta z -\frac{\sin\phi}{x},\label{eq:dphi/ds}
\end{align}
\begin{align}
    \frac{\mathrm{d}x}{\mathrm{d}s} = \cos\phi,\label{eq:dx/ds}
\end{align}
\begin{align}
    \frac{\mathrm{d}z}{\mathrm{d}s} = \sin\phi,\label{eq:dz/ds}
\end{align}
where $\beta = \rho g/\gamma$. The shape of the droplet can be calculated by integrating Eqs. \eqref{eq:dphi/ds}--\eqref{eq:dz/ds} by $s$ from $s=0$ with $\phi(s = 0) =0$. By comparing the calculated curve with the image recorded in the experiments we obtained $b$, $\beta$, and thus,the interfacial tension $\gamma$ using $\gamma = \rho g/\beta$.

\subsection{Analysis method}
We recorded a sequence of droplet images every 60~s and calculated the interfacial tension for each image in the experiments. The droplet shape is described in the reference frame illustrated in Fig.~\ref{fig:coordinate}(a). The vertical direction of images obtained in an experiment can differ from the actual vertical direction since the camera can be set at a slightly rotated angle. To compensate for the effect, we rotated the images with an angle $\theta$ for the most symmetric distribution of the image with respect to the vertical line. To find $\theta$, the value of:
\begin{align}
    E_{\theta} =\sum_z \left(X_r(z,\theta)+X_l(z,\theta)\right)^2
\end{align}
was minimized. Here, $X_r(z,\theta)$ $(>0)$ and $X_l(z,\theta)$ $(<0)$ are the $x$ coordinates of the right and left peripheries at height $z$ for the rotated image. Since the position of the camera was fixed, $\theta$ should be identical for all the images of the sequence. Thus, we adopted the average of $\theta$ obtained by minimizing $E_{\theta}$ using the data of extracted 10 images. In the following, we use the rotated image, and the top point of the droplet estimated from the image is set to be the coordinate origin.

Next, the transformed data are compared with numerical results obtained from Eqs. \eqref{eq:dphi/ds}--\eqref{eq:dz/ds}. This calculation yields a set of coordinates $(x_i, z_i)$ along the edge of the droplet depending on the parameters $b$ and $\beta$ as well as on the  coordinates of the top point $x_{\mathrm{tp}}$ and $z_{\mathrm{tp}}$.

Then, we compared the calculated coordinates at the droplet periphery $(x_i, z_i)$ and that obtained by the experiments $(X_j, Z_j)$. For the points $(X_j, Z_j)$ close to the origin, i.e.,  $\phi<\pi/4$, we compared the value of $z$ for each $x$. For the points $(X_i, Z_i)$ below them, i.e., $\phi>\pi/4$, we compared the value of $x$ for each $z$. Since the error in the region close to the needle should be large, we excluded the data where the width of the droplet was less than 1.5 times of the thickness of the needle. Then, we minimized
\begin{align}
    E(\beta, b, x_{\mathrm{tp}}, z_{\mathrm{tp}}) =& \sum_i\left(X_i-x_i-x_{\mathrm{tp}}\right)^2 \nonumber \\
    &+ \sum_j\left(Z_j-z_j-z_{\mathrm{tp}}\right)^2,
\end{align}
with respect to $b$, $\beta$, $x_{\mathrm{tp}}$ and $z_{\mathrm{tp}}$. The top point of the droplet ($x_{\mathrm{tp}}$, $z_{\mathrm{tp}}$) should roughly match the origin, though it can differ because of the finite size of the pixel. $x_{\mathrm{tp}}$ and $z_{\mathrm{tp}}$ are introduced to compensate this effect. From the value of $\beta$, we get $\gamma_{\mathrm{ow}}$. It should be noted that the initial values for the parameters to minimize $E(\beta,b,x_{\mathrm{tp}},z_{\mathrm{tp}})$ are determined as follows; For the first image ($t=0$), we estimate the interfacial tension $\gamma_0$ from the maximum width $D$ of the droplet, and the width $d$ at $z=D$ using $d/D$ method\cite{doi:10.1098/rspa.1948.0063} and adopt $\beta=\rho g/\gamma_0$, $b=D/2$ and $x_{\mathrm{tp}}=z_{\mathrm{tp}}=0$ as the initial values of the parameters. For the second or later image, we use the values of $\beta$, $b$, $x_{\mathrm{tp}}$, and $z_{\mathrm{tp}}$ that minimize $E(\beta,b,x_{\mathrm{tp}},z_{\mathrm{tp}})$ in the previous image as the initial values to accelerate the calculation.

\subsection{\label{appsub:line_tension}Derivation of the expression for line tension}

Assuming the characteristic length of deformation of a triple line is much larger than the capillary length, we consider the deformation of the interface near the straight triple line. We set the Cartesian coordinates so that the $y$-axis matches the triple line and $z = \zeta_{\mathrm{ao}}(x)\, (x>0)$ and $z = -\zeta_{\mathrm{ow}}(x)\, (x>0)$ indicate the interface between air and oil and that between oil and water, respectively, as shown in Fig.~\ref{fig:tripleline}. Then, the excess energy of the triple line per unit length is described as
\begin{align}
    \mathcal{T} =\mathcal{T}_{\mathrm{ao}} + \mathcal{T}_{\mathrm{ow}},
\end{align}
where $\mathcal{T}_{\mathrm{ao}}$ and $\mathcal{T}_{\mathrm{ow}}$ originate from the air/oil and oil/water interfaces, respectively, and they are explicitly estimated as\cite{Gennes2004CapillarityAW}
\begin{align}
    \mathcal{T}_{\mathrm{ao}}
    =&\int_0^{\infty}\mathrm{d}x\left[
    \gamma_{\mathrm{ao}}\left(\frac{1}{\cos\theta_{\mathrm{ao}}(x)}-1\right) \right. \nonumber \\
    & \qquad \left. +\frac{1}{2}\rho_{\mathrm{ao}}g(e_{\mathrm{ao}}-\zeta_{\mathrm{ao}}(x))^2
    \right],
\end{align}
\begin{align}
    \mathcal{T}_{\mathrm{ow}}
    =& \int_0^{\infty}\mathrm{d}x\left[
    \gamma_{\mathrm{ow}}\left(\frac{1}{\cos\theta_{\mathrm{ow}}(x)}-1\right) \right. \nonumber \\
    & \qquad + \left.\frac{1}{2}\rho_{\mathrm{ow}}g(e_{\mathrm{ow}}-\zeta_{\mathrm{ow}}(x))^2
    \right],
\end{align}
where 
 $\tan\theta_{\mathrm{ao}}(x)=\mathrm{d}\zeta_{\mathrm{ao}}/\mathrm{d}x$ and $\tan\theta_{\mathrm{ow}}(x)=\mathrm{d}\zeta_{\mathrm{ow}}/\mathrm{d}x$. The shape of the droplet $\zeta_{\mathrm{ao}}(x)$ and $\zeta_{\mathrm{ow}}(x)$ can be calculated by minimizing $\mathcal{T}_{\mathrm{ao}}$ and $\mathcal{T}_{\mathrm{ow}}$ with respect to $\zeta_{\mathrm{ao}}$ and $\zeta_{\mathrm{ow}}$. Then, we get
\begin{align}
    -\gamma_{\mathrm{ao}}\cos^3\theta_{\mathrm{ao}}(x)\frac{\mathrm{d}^2\zeta_{\mathrm{ao}}}{\mathrm{d}x^2} = \rho_{\mathrm{ao}}g(e_{\mathrm{ao}}-\zeta_{\mathrm{ao}}(x)),
\end{align}
\begin{align}
    -\gamma_{\mathrm{ow}}\cos^3\theta_{\mathrm{ow}}(x)\frac{\mathrm{d}^2\zeta_{\mathrm{ow}}}{\mathrm{d}x^2} = \rho_{\mathrm{ow}}g(e_{\mathrm{ow}}-\zeta_{\mathrm{ow}}(x)),
\end{align}
which lead
\begin{align}
    \mathcal{T}_{\mathrm{ao}}  =\frac{4}{3}\gamma_{\mathrm{ao}}\kappa_{\mathrm{ao}}^{-1}\left[1-\cos^2\left(\frac{\theta_{\mathrm{ao}}(x=0)}{2}\right)\right],
\end{align}
\begin{align}
    \mathcal{T}_{\mathrm{ow}} =\frac{4}{3}\gamma_{\mathrm{ow}}\kappa_{\mathrm{ow}}^{-1}\left[1-\cos^2\left(\frac{\theta_{\mathrm{ow}}(x=0)}{2}\right)\right],
\end{align}
where $\kappa_{\mathrm{ao}}^{-1}$ and $\kappa_{\mathrm{ow}}^{-1}$ are capillary lengths of air/oil and oil/water interfaces, respectively. From these equations, we derived Eq.~\eqref{eq:line_tension}.

Assuming that the water surface outside of the droplet is completely flat, we obtain
\begin{align}
    \cos \left(\theta_{\mathrm{ao}}(x=0)\right)
    &=\frac{\gamma_{\mathrm{aw}}^2-\gamma_{\mathrm{ow}}^2+\gamma_{\mathrm{ao}}^2}{2\gamma_{\mathrm{aw}}\gamma_{\mathrm{ao}}},
\end{align}
\begin{align}
    \cos \left(\theta_{\mathrm{ow}}(x=0)\right)
    &=\frac{\gamma_{\mathrm{aw}}^2-\gamma_{\mathrm{ao}}^2+\gamma_{\mathrm{ow}}^2}{2\gamma_{\mathrm{aw}}\gamma_{\mathrm{ow}}},
\end{align}
which relates to the excess energy of the triple line with the values of interfacial tensions.

\section{\label{app:QELS}Water surface tension measurement around a droplet}

The surface tension around a $50~\mu\mathrm{L}$ paraffin droplet with camphor (concentration $0.06~\mathrm{M}$) and oil red O (concentration $1.0~\mathrm{g/L}$) was measured using quasi-elastic laser scattering (QELS) method\cite{doi:10.1021/acs.jpcb.3c00466}. We put a metal screw on the bottom of the Petri dish upside down and then placed a droplet on the tip of the screw so that the droplet was pinned. We measured the surface tension at the $5~\mathrm{mm}$ distance from the droplet. The optical settings for the surface tension measurement were the same as described in  \cite{KARASAWA2018184,doi:10.1246/cl.140201}.

\begin{figure}[tb]
    \centering
    \includegraphics{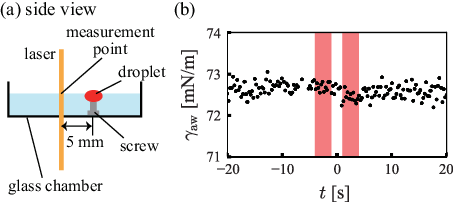}
    \caption{\label{fig:QELS}
   (a) Geometry of experimental setup. (b) Surface tension as a function of time $t$ after locating a droplet on the water surface. The red shaded regions mark time intervals $-4~\mathrm{s}<t<-1~\mathrm{s}$ and $1~\mathrm{s}<t<4~\mathrm{s}$. The paraffin droplet included camphor at $0.06~\mathrm{M}$ and oil red O at $1.0~\mathrm{g/L}$.}
\end{figure}

The surface tension $\gamma_\mathrm{aw}$ as the function of time $t$ is plotted in Fig.~\ref{fig:QELS}. Time $t=0$ corresponds to the moment when the droplet was placed on the tip of the screw. The average values of the surface tension $\gamma_\mathrm{aw}$ were $72.70~\mathrm{mN/m}$ at $-4~\mathrm{s}<t<-1~\mathrm{s}$ and $72.45~\mathrm{mN/m}$ at $1~\mathrm{s}<t<4~\mathrm{s}$. Using these values, we estimate $\delta\gamma_\mathrm{aw}$ as $0.25~\mathrm{mN/m}$ for the discussion in subsection \ref{subsec:discuttion1}.

\begin{thebibliography}{48}%
\makeatletter
\providecommand \@ifxundefined [1]{%
 \@ifx{#1\undefined}
}%
\providecommand \@ifnum [1]{%
 \ifnum #1\expandafter \@firstoftwo
 \else \expandafter \@secondoftwo
 \fi
}%
\providecommand \@ifx [1]{%
 \ifx #1\expandafter \@firstoftwo
 \else \expandafter \@secondoftwo
 \fi
}%
\providecommand \natexlab [1]{#1}%
\providecommand \enquote  [1]{``#1''}%
\providecommand \bibnamefont  [1]{#1}%
\providecommand \bibfnamefont [1]{#1}%
\providecommand \citenamefont [1]{#1}%
\providecommand \href@noop [0]{\@secondoftwo}%
\providecommand \href [0]{\begingroup \@sanitize@url \@href}%
\providecommand \@href[1]{\@@startlink{#1}\@@href}%
\providecommand \@@href[1]{\endgroup#1\@@endlink}%
\providecommand \@sanitize@url [0]{\catcode `\\12\catcode `\$12\catcode
  `\&12\catcode `\#12\catcode `\^12\catcode `\_12\catcode `\%12\relax}%
\providecommand \@@startlink[1]{}%
\providecommand \@@endlink[0]{}%
\providecommand \url  [0]{\begingroup\@sanitize@url \@url }%
\providecommand \@url [1]{\endgroup\@href {#1}{\urlprefix }}%
\providecommand \urlprefix  [0]{URL }%
\providecommand \Eprint [0]{\href }%
\providecommand \doibase [0]{https://doi.org/}%
\providecommand \selectlanguage [0]{\@gobble}%
\providecommand \bibinfo  [0]{\@secondoftwo}%
\providecommand \bibfield  [0]{\@secondoftwo}%
\providecommand \translation [1]{[#1]}%
\providecommand \BibitemOpen [0]{}%
\providecommand \bibitemStop [0]{}%
\providecommand \bibitemNoStop [0]{.\EOS\space}%
\providecommand \EOS [0]{\spacefactor3000\relax}%
\providecommand \BibitemShut  [1]{\csname bibitem#1\endcsname}%
\let\auto@bib@innerbib\@empty
\bibitem [{\citenamefont {Luisi}(2006)}]{Luisi}%
  \BibitemOpen
  \bibfield  {author} {\bibinfo {author} {\bibfnamefont {P.}~\bibnamefont
  {Luisi}},\ }\href {https://books.google.co.jp/books?id=1Oxfq5VTcDkC} {\emph
  {\bibinfo {title} {The Emergence of Life: From Chemical Origins to Synthetic
  Biology}}}\ (\bibinfo  {publisher} {Cambridge University Press},\ \bibinfo
  {year} {2006})\BibitemShut {NoStop}%
\bibitem [{\citenamefont {Nakata}\ \emph {et~al.}(2018)\citenamefont {Nakata},
  \citenamefont {Pimienta}, \citenamefont {Lagzi}, \citenamefont {Kitahata},\
  and\ \citenamefont {Suematsu}}]{10.1039/9781788013499}%
  \BibitemOpen
  \bibfield  {author} {\bibinfo {author} {\bibfnamefont {S.}~\bibnamefont
  {Nakata}}, \bibinfo {author} {\bibfnamefont {V.}~\bibnamefont {Pimienta}},
  \bibinfo {author} {\bibfnamefont {I.}~\bibnamefont {Lagzi}}, \bibinfo
  {author} {\bibfnamefont {H.}~\bibnamefont {Kitahata}},\ and\ \bibinfo
  {author} {\bibfnamefont {N.~J.}\ \bibnamefont {Suematsu}},\ }\href
  {https://doi.org/10.1039/9781788013499} {\emph {\bibinfo {title}
  {Self-organized Motion: Physicochemical Design based on Nonlinear
  Dynamics}}}\ (\bibinfo  {publisher} {The Royal Society of Chemistry},\
  \bibinfo {year} {2018})\BibitemShut {NoStop}%
\bibitem [{\citenamefont {Nakata}\ \emph {et~al.}(2015)\citenamefont {Nakata},
  \citenamefont {Nagayama}, \citenamefont {Kitahata}, \citenamefont
  {Suematsu},\ and\ \citenamefont {Hasegawa}}]{C5CP00541H}%
  \BibitemOpen
  \bibfield  {author} {\bibinfo {author} {\bibfnamefont {S.}~\bibnamefont
  {Nakata}}, \bibinfo {author} {\bibfnamefont {M.}~\bibnamefont {Nagayama}},
  \bibinfo {author} {\bibfnamefont {H.}~\bibnamefont {Kitahata}}, \bibinfo
  {author} {\bibfnamefont {N.~J.}\ \bibnamefont {Suematsu}},\ and\ \bibinfo
  {author} {\bibfnamefont {T.}~\bibnamefont {Hasegawa}},\ }\href
  {https://doi.org/10.1039/C5CP00541H} {\bibfield  {journal} {\bibinfo
  {journal} {Phys. Chem. Chem. Phys.}\ }\textbf {\bibinfo {volume} {17}},\
  \bibinfo {pages} {10326} (\bibinfo {year} {2015})}\BibitemShut {NoStop}%
\bibitem [{\citenamefont {Tomlinson}\ and\ \citenamefont
  {Miller}(1862)}]{doi:10.1098/rspl.1860.0124}%
  \BibitemOpen
  \bibfield  {author} {\bibinfo {author} {\bibfnamefont {C.}~\bibnamefont
  {Tomlinson}}\ and\ \bibinfo {author} {\bibfnamefont {W.~A.}\ \bibnamefont
  {Miller}},\ }\href {https://doi.org/10.1098/rspl.1860.0124} {\bibfield
  {journal} {\bibinfo  {journal} {Proc. R. Soc. London}\ }\textbf {\bibinfo
  {volume} {11}},\ \bibinfo {pages} {575} (\bibinfo {year} {1862})}\BibitemShut
  {NoStop}%
\bibitem [{\citenamefont {Strutt}(1890)}]{doi:10.1098/rspl.1889.0099}%
  \BibitemOpen
  \bibfield  {author} {\bibinfo {author} {\bibfnamefont {R.~J.}\ \bibnamefont
  {Strutt}},\ }\href {https://doi.org/10.1098/rspl.1889.0099} {\bibfield
  {journal} {\bibinfo  {journal} {Proc. R. Soc. London}\ }\textbf {\bibinfo
  {volume} {47}},\ \bibinfo {pages} {364} (\bibinfo {year} {1890})}\BibitemShut
  {NoStop}%
\bibitem [{\citenamefont {Karasawa}\ \emph {et~al.}(2018)\citenamefont
  {Karasawa}, \citenamefont {Nomoto}, \citenamefont {Chiari}, \citenamefont
  {Toyota},\ and\ \citenamefont {Fujinami}}]{KARASAWA2018184}%
  \BibitemOpen
  \bibfield  {author} {\bibinfo {author} {\bibfnamefont {Y.}~\bibnamefont
  {Karasawa}}, \bibinfo {author} {\bibfnamefont {T.}~\bibnamefont {Nomoto}},
  \bibinfo {author} {\bibfnamefont {L.}~\bibnamefont {Chiari}}, \bibinfo
  {author} {\bibfnamefont {T.}~\bibnamefont {Toyota}},\ and\ \bibinfo {author}
  {\bibfnamefont {M.}~\bibnamefont {Fujinami}},\ }\href
  {https://doi.org/https://doi.org/10.1016/j.jcis.2017.09.099} {\bibfield
  {journal} {\bibinfo  {journal} {J. Colloid Interface Sci.}\ }\textbf
  {\bibinfo {volume} {511}},\ \bibinfo {pages} {184} (\bibinfo {year}
  {2018})}\BibitemShut {NoStop}%
\bibitem [{\citenamefont {Karasawa}\ \emph {et~al.}(2014)\citenamefont
  {Karasawa}, \citenamefont {Oshima}, \citenamefont {Nomoto}, \citenamefont
  {Toyota},\ and\ \citenamefont {Fujinami}}]{doi:10.1246/cl.140201}%
  \BibitemOpen
  \bibfield  {author} {\bibinfo {author} {\bibfnamefont {Y.}~\bibnamefont
  {Karasawa}}, \bibinfo {author} {\bibfnamefont {S.}~\bibnamefont {Oshima}},
  \bibinfo {author} {\bibfnamefont {T.}~\bibnamefont {Nomoto}}, \bibinfo
  {author} {\bibfnamefont {T.}~\bibnamefont {Toyota}},\ and\ \bibinfo {author}
  {\bibfnamefont {M.}~\bibnamefont {Fujinami}},\ }\href
  {https://doi.org/10.1246/cl.140201} {\bibfield  {journal} {\bibinfo
  {journal} {Chem. Lett.}\ }\textbf {\bibinfo {volume} {43}},\ \bibinfo {pages}
  {1002} (\bibinfo {year} {2014})}\BibitemShut {NoStop}%
\bibitem [{\citenamefont {Nomoto}\ \emph {et~al.}(2023)\citenamefont {Nomoto},
  \citenamefont {Marumo}, \citenamefont {Chiari}, \citenamefont {Toyota},\ and\
  \citenamefont {Fujinami}}]{doi:10.1021/acs.jpcb.3c00466}%
  \BibitemOpen
  \bibfield  {author} {\bibinfo {author} {\bibfnamefont {T.}~\bibnamefont
  {Nomoto}}, \bibinfo {author} {\bibfnamefont {M.}~\bibnamefont {Marumo}},
  \bibinfo {author} {\bibfnamefont {L.}~\bibnamefont {Chiari}}, \bibinfo
  {author} {\bibfnamefont {T.}~\bibnamefont {Toyota}},\ and\ \bibinfo {author}
  {\bibfnamefont {M.}~\bibnamefont {Fujinami}},\ }\href
  {https://doi.org/10.1021/acs.jpcb.3c00466} {\bibfield  {journal} {\bibinfo
  {journal} {J. Phys. Chem. B}\ }\textbf {\bibinfo {volume} {127}},\ \bibinfo
  {pages} {2863} (\bibinfo {year} {2023})}\BibitemShut {NoStop}%
\bibitem [{\citenamefont {Nagayama}\ \emph {et~al.}(2004)\citenamefont
  {Nagayama}, \citenamefont {Nakata}, \citenamefont {Doi},\ and\ \citenamefont
  {Hayashima}}]{NAGAYAMA2004151}%
  \BibitemOpen
  \bibfield  {author} {\bibinfo {author} {\bibfnamefont {M.}~\bibnamefont
  {Nagayama}}, \bibinfo {author} {\bibfnamefont {S.}~\bibnamefont {Nakata}},
  \bibinfo {author} {\bibfnamefont {Y.}~\bibnamefont {Doi}},\ and\ \bibinfo
  {author} {\bibfnamefont {Y.}~\bibnamefont {Hayashima}},\ }\href
  {https://doi.org/https://doi.org/10.1016/j.physd.2004.02.003} {\bibfield
  {journal} {\bibinfo  {journal} {Physica D: Nonlinear Phenomena}\ }\textbf
  {\bibinfo {volume} {194}},\ \bibinfo {pages} {151} (\bibinfo {year}
  {2004})}\BibitemShut {NoStop}%
\bibitem [{\citenamefont {Hayashima}\ \emph {et~al.}(2001)\citenamefont
  {Hayashima}, \citenamefont {Nagayama},\ and\ \citenamefont
  {Nakata}}]{doi:10.1021/jp004505n}%
  \BibitemOpen
  \bibfield  {author} {\bibinfo {author} {\bibfnamefont {Y.}~\bibnamefont
  {Hayashima}}, \bibinfo {author} {\bibfnamefont {M.}~\bibnamefont
  {Nagayama}},\ and\ \bibinfo {author} {\bibfnamefont {S.}~\bibnamefont
  {Nakata}},\ }\href {https://doi.org/10.1021/jp004505n} {\bibfield  {journal}
  {\bibinfo  {journal} {J. Phys. Chem. B}\ }\textbf {\bibinfo {volume} {105}},\
  \bibinfo {pages} {5353} (\bibinfo {year} {2001})}\BibitemShut {NoStop}%
\bibitem [{\citenamefont {Nakata}\ \emph {et~al.}(1997)\citenamefont {Nakata},
  \citenamefont {Iguchi}, \citenamefont {Ose}, \citenamefont {Kuboyama},
  \citenamefont {Ishii},\ and\ \citenamefont
  {Yoshikawa}}]{doi:10.1021/la970196p}%
  \BibitemOpen
  \bibfield  {author} {\bibinfo {author} {\bibfnamefont {S.}~\bibnamefont
  {Nakata}}, \bibinfo {author} {\bibfnamefont {Y.}~\bibnamefont {Iguchi}},
  \bibinfo {author} {\bibfnamefont {S.}~\bibnamefont {Ose}}, \bibinfo {author}
  {\bibfnamefont {M.}~\bibnamefont {Kuboyama}}, \bibinfo {author}
  {\bibfnamefont {T.}~\bibnamefont {Ishii}},\ and\ \bibinfo {author}
  {\bibfnamefont {K.}~\bibnamefont {Yoshikawa}},\ }\href
  {https://doi.org/10.1021/la970196p} {\bibfield  {journal} {\bibinfo
  {journal} {Langmuir}\ }\textbf {\bibinfo {volume} {13}},\ \bibinfo {pages}
  {4454} (\bibinfo {year} {1997})}\BibitemShut {NoStop}%
\bibitem [{\citenamefont {Boniface}\ \emph {et~al.}(2019)\citenamefont
  {Boniface}, \citenamefont {Cottin-Bizonne}, \citenamefont {Kervil},
  \citenamefont {Ybert},\ and\ \citenamefont
  {Detcheverry}}]{PhysRevE.99.062605}%
  \BibitemOpen
  \bibfield  {author} {\bibinfo {author} {\bibfnamefont {D.}~\bibnamefont
  {Boniface}}, \bibinfo {author} {\bibfnamefont {C.}~\bibnamefont
  {Cottin-Bizonne}}, \bibinfo {author} {\bibfnamefont {R.}~\bibnamefont
  {Kervil}}, \bibinfo {author} {\bibfnamefont {C.}~\bibnamefont {Ybert}},\ and\
  \bibinfo {author} {\bibfnamefont {F.}~\bibnamefont {Detcheverry}},\ }\href
  {https://doi.org/10.1103/PhysRevE.99.062605} {\bibfield  {journal} {\bibinfo
  {journal} {Phys. Rev. E}\ }\textbf {\bibinfo {volume} {99}},\ \bibinfo
  {pages} {062605} (\bibinfo {year} {2019})}\BibitemShut {NoStop}%
\bibitem [{\citenamefont {Morohashi}\ \emph {et~al.}(2019)\citenamefont
  {Morohashi}, \citenamefont {Imai},\ and\ \citenamefont
  {Toyota}}]{MOROHASHI2019104}%
  \BibitemOpen
  \bibfield  {author} {\bibinfo {author} {\bibfnamefont {H.}~\bibnamefont
  {Morohashi}}, \bibinfo {author} {\bibfnamefont {M.}~\bibnamefont {Imai}},\
  and\ \bibinfo {author} {\bibfnamefont {T.}~\bibnamefont {Toyota}},\ }\href
  {https://doi.org/https://doi.org/10.1016/j.cplett.2019.02.034} {\bibfield
  {journal} {\bibinfo  {journal} {Chem. Phys. Lett.}\ }\textbf {\bibinfo
  {volume} {721}},\ \bibinfo {pages} {104} (\bibinfo {year}
  {2019})}\BibitemShut {NoStop}%
\bibitem [{\citenamefont {Tiwari}\ \emph {et~al.}(2020)\citenamefont {Tiwari},
  \citenamefont {Parmananda},\ and\ \citenamefont {Chelakkot}}]{D0SM01393E}%
  \BibitemOpen
  \bibfield  {author} {\bibinfo {author} {\bibfnamefont {I.}~\bibnamefont
  {Tiwari}}, \bibinfo {author} {\bibfnamefont {P.}~\bibnamefont {Parmananda}},\
  and\ \bibinfo {author} {\bibfnamefont {R.}~\bibnamefont {Chelakkot}},\ }\href
  {https://doi.org/10.1039/D0SM01393E} {\bibfield  {journal} {\bibinfo
  {journal} {Soft Matter}\ }\textbf {\bibinfo {volume} {16}},\ \bibinfo {pages}
  {10334} (\bibinfo {year} {2020})}\BibitemShut {NoStop}%
\bibitem [{\citenamefont {Kitahata}\ and\ \citenamefont
  {Koyano}(2020)}]{doi:10.7566/JPSJ.89.094001}%
  \BibitemOpen
  \bibfield  {author} {\bibinfo {author} {\bibfnamefont {H.}~\bibnamefont
  {Kitahata}}\ and\ \bibinfo {author} {\bibfnamefont {Y.}~\bibnamefont
  {Koyano}},\ }\href {https://doi.org/10.7566/JPSJ.89.094001} {\bibfield
  {journal} {\bibinfo  {journal} {J. Phys. Soc. Jpn.}\ }\textbf {\bibinfo
  {volume} {89}},\ \bibinfo {pages} {094001} (\bibinfo {year}
  {2020})}\BibitemShut {NoStop}%
\bibitem [{\citenamefont {Kitahata}\ and\ \citenamefont
  {Koyano}(2022)}]{10.3389/fphy.2022.858791}%
  \BibitemOpen
  \bibfield  {author} {\bibinfo {author} {\bibfnamefont {H.}~\bibnamefont
  {Kitahata}}\ and\ \bibinfo {author} {\bibfnamefont {Y.}~\bibnamefont
  {Koyano}},\ }\href
  {https://www.frontiersin.org/articles/10.3389/fphy.2022.858791} {\bibfield
  {journal} {\bibinfo  {journal} {Front. Phys.}\ }\textbf {\bibinfo {volume}
  {10}},\  \bibinfo {pages} {858791} (\bibinfo {year} {2022})}\BibitemShut {NoStop}%
\bibitem [{\citenamefont {Kitahata}\ \emph {et~al.}(2013)\citenamefont
  {Kitahata}, \citenamefont {Iida},\ and\ \citenamefont
  {Nagayama}}]{PhysRevE.87.010901}%
  \BibitemOpen
  \bibfield  {author} {\bibinfo {author} {\bibfnamefont {H.}~\bibnamefont
  {Kitahata}}, \bibinfo {author} {\bibfnamefont {K.}~\bibnamefont {Iida}},\
  and\ \bibinfo {author} {\bibfnamefont {M.}~\bibnamefont {Nagayama}},\ }\href
  {https://doi.org/10.1103/PhysRevE.87.010901} {\bibfield  {journal} {\bibinfo
  {journal} {Phys. Rev. E}\ }\textbf {\bibinfo {volume} {87}},\ \bibinfo
  {pages} {010901} (\bibinfo {year} {2013})}\BibitemShut {NoStop}%
\bibitem [{\citenamefont {Iida}\ \emph {et~al.}(2014)\citenamefont {Iida},
  \citenamefont {Kitahata},\ and\ \citenamefont
  {Nagayama}}]{iida2014theoretical}%
  \BibitemOpen
  \bibfield  {author} {\bibinfo {author} {\bibfnamefont {K.}~\bibnamefont
  {Iida}}, \bibinfo {author} {\bibfnamefont {H.}~\bibnamefont {Kitahata}},\
  and\ \bibinfo {author} {\bibfnamefont {M.}~\bibnamefont {Nagayama}},\
  }\href@noop {} {\bibfield  {journal} {\bibinfo  {journal} {Physica D:
  Nonlinear Phenomena}\ }\textbf {\bibinfo {volume} {272}},\ \bibinfo {pages}
  {39} (\bibinfo {year} {2014})}\BibitemShut {NoStop}%
\bibitem [{\citenamefont {Bassik}\ \emph {et~al.}(2008)\citenamefont {Bassik},
  \citenamefont {Abebe},\ and\ \citenamefont
  {Gracias}}]{doi:10.1021/la801329g}%
  \BibitemOpen
  \bibfield  {author} {\bibinfo {author} {\bibfnamefont {N.}~\bibnamefont
  {Bassik}}, \bibinfo {author} {\bibfnamefont {B.~T.}\ \bibnamefont {Abebe}},\
  and\ \bibinfo {author} {\bibfnamefont {D.~H.}\ \bibnamefont {Gracias}},\
  }\href {https://doi.org/10.1021/la801329g} {\bibfield  {journal} {\bibinfo
  {journal} {Langmuir}\ }\textbf {\bibinfo {volume} {24}},\ \bibinfo {pages}
  {12158} (\bibinfo {year} {2008})}\BibitemShut {NoStop}%
\bibitem [{\citenamefont {Akella}\ \emph {et~al.}(2018)\citenamefont {Akella},
  \citenamefont {Singh}, \citenamefont {Mandre},\ and\ \citenamefont
  {Bandi}}]{AKELLA20181176}%
  \BibitemOpen
  \bibfield  {author} {\bibinfo {author} {\bibfnamefont {V.}~\bibnamefont
  {Akella}}, \bibinfo {author} {\bibfnamefont {D.~K.}\ \bibnamefont {Singh}},
  \bibinfo {author} {\bibfnamefont {S.}~\bibnamefont {Mandre}},\ and\ \bibinfo
  {author} {\bibfnamefont {M.}~\bibnamefont {Bandi}},\ }\href
  {https://doi.org/https://doi.org/10.1016/j.physleta.2018.02.026} {\bibfield
  {journal} {\bibinfo  {journal} {Phys. Lett. A}\ }\textbf {\bibinfo {volume}
  {382}},\ \bibinfo {pages} {1176} (\bibinfo {year} {2018})}\BibitemShut
  {NoStop}%
\bibitem [{\citenamefont {B\'{a}ns\'{a}gi}\ \emph {et~al.}(2013)\citenamefont
  {B\'{a}ns\'{a}gi}, \citenamefont {Wrobel}, \citenamefont {Scott},\ and\
  \citenamefont {Taylor}}]{doi:10.1021/jp405364c}%
  \BibitemOpen
  \bibfield  {author} {\bibinfo {author} {\bibfnamefont {T.~J.}\ \bibnamefont
  {B\'{a}ns\'{a}gi}}, \bibinfo {author} {\bibfnamefont {M.~M.}\ \bibnamefont
  {Wrobel}}, \bibinfo {author} {\bibfnamefont {S.~K.}\ \bibnamefont {Scott}},\
  and\ \bibinfo {author} {\bibfnamefont {A.~F.}\ \bibnamefont {Taylor}},\
  }\href {https://doi.org/10.1021/jp405364c} {\bibfield  {journal} {\bibinfo
  {journal} {J. Phys. Chem. B}\ }\textbf {\bibinfo {volume} {117}},\ \bibinfo
  {pages} {13572} (\bibinfo {year} {2013})}\BibitemShut {NoStop}%
\bibitem [{\citenamefont {L\"{o}ffler}\ \emph {et~al.}(2021)\citenamefont
  {L\"{o}ffler}, \citenamefont {Hanczyc},\ and\ \citenamefont
  {Gorecki}}]{molecules26113116}%
  \BibitemOpen
  \bibfield  {author} {\bibinfo {author} {\bibfnamefont {R.~J.~G.}\
  \bibnamefont {L\"{o}ffler}}, \bibinfo {author} {\bibfnamefont {M.~M.}\
  \bibnamefont {Hanczyc}},\ and\ \bibinfo {author} {\bibfnamefont
  {J.}~\bibnamefont {Gorecki}},\ }\href
  {https://www.mdpi.com/1420-3049/26/11/3116} {\bibfield  {journal} {\bibinfo
  {journal} {Molecules}\ }\textbf {\bibinfo {volume} {26}} (\bibinfo {year}
  {2021})}\BibitemShut {NoStop}%
\bibitem [{\citenamefont {Magome}\ and\ \citenamefont
  {Yoshikawa}(1996)}]{doi:10.1021/jp9616876}%
  \BibitemOpen
  \bibfield  {author} {\bibinfo {author} {\bibfnamefont {N.}~\bibnamefont
  {Magome}}\ and\ \bibinfo {author} {\bibfnamefont {K.}~\bibnamefont
  {Yoshikawa}},\ }\href {https://doi.org/10.1021/jp9616876} {\bibfield
  {journal} {\bibinfo  {journal} {J. Phys. Chem.}\ }\textbf {\bibinfo {volume}
  {100}},\ \bibinfo {pages} {19102} (\bibinfo {year} {1996})}\BibitemShut
  {NoStop}%
\bibitem [{\citenamefont {Sumino}\ \emph {et~al.}(2005)\citenamefont {Sumino},
  \citenamefont {Magome}, \citenamefont {Hamada},\ and\ \citenamefont
  {Yoshikawa}}]{PhysRevLett.94.068301}%
  \BibitemOpen
  \bibfield  {author} {\bibinfo {author} {\bibfnamefont {Y.}~\bibnamefont
  {Sumino}}, \bibinfo {author} {\bibfnamefont {N.}~\bibnamefont {Magome}},
  \bibinfo {author} {\bibfnamefont {T.}~\bibnamefont {Hamada}},\ and\ \bibinfo
  {author} {\bibfnamefont {K.}~\bibnamefont {Yoshikawa}},\ }\href
  {https://doi.org/10.1103/PhysRevLett.94.068301} {\bibfield  {journal}
  {\bibinfo  {journal} {Phys. Rev. Lett.}\ }\textbf {\bibinfo {volume} {94}},\
  \bibinfo {pages} {068301} (\bibinfo {year} {2005})}\BibitemShut {NoStop}%
\bibitem [{\citenamefont {Nagai}\ \emph {et~al.}(2005)\citenamefont {Nagai},
  \citenamefont {Sumino}, \citenamefont {Kitahata},\ and\ \citenamefont
  {Yoshikawa}}]{PhysRevE.71.065301}%
  \BibitemOpen
  \bibfield  {author} {\bibinfo {author} {\bibfnamefont {K.}~\bibnamefont
  {Nagai}}, \bibinfo {author} {\bibfnamefont {Y.}~\bibnamefont {Sumino}},
  \bibinfo {author} {\bibfnamefont {H.}~\bibnamefont {Kitahata}},\ and\
  \bibinfo {author} {\bibfnamefont {K.}~\bibnamefont {Yoshikawa}},\ }\href
  {https://doi.org/10.1103/PhysRevE.71.065301} {\bibfield  {journal} {\bibinfo
  {journal} {Phys. Rev. E}\ }\textbf {\bibinfo {volume} {71}},\ \bibinfo
  {pages} {065301} (\bibinfo {year} {2005})}\BibitemShut {NoStop}%
\bibitem [{\citenamefont {Nagai}\ \emph {et~al.}(2006)\citenamefont {Nagai},
  \citenamefont {Sumino}, \citenamefont {Kitahata},\ and\ \citenamefont
  {Yoshikawa}}]{10.1143/PTPS.161.286}%
  \BibitemOpen
  \bibfield  {author} {\bibinfo {author} {\bibfnamefont {K.}~\bibnamefont
  {Nagai}}, \bibinfo {author} {\bibfnamefont {Y.}~\bibnamefont {Sumino}},
  \bibinfo {author} {\bibfnamefont {H.}~\bibnamefont {Kitahata}},\ and\
  \bibinfo {author} {\bibfnamefont {K.}~\bibnamefont {Yoshikawa}},\ }\href
  {https://doi.org/10.1143/PTPS.161.286} {\bibfield  {journal} {\bibinfo
  {journal} {Prog. Theor. Phys. Suppl.}\ }\textbf {\bibinfo {volume} {161}},\
  \bibinfo {pages} {286} (\bibinfo {year} {2006})}\BibitemShut {NoStop}%
\bibitem [{\citenamefont {Boniface}\ \emph {et~al.}(2022)\citenamefont
  {Boniface}, \citenamefont {Sebilleau}, \citenamefont {Magnaudet},\ and\
  \citenamefont {Pimienta}}]{BONIFACE2022990}%
  \BibitemOpen
  \bibfield  {author} {\bibinfo {author} {\bibfnamefont {D.}~\bibnamefont
  {Boniface}}, \bibinfo {author} {\bibfnamefont {J.}~\bibnamefont {Sebilleau}},
  \bibinfo {author} {\bibfnamefont {J.}~\bibnamefont {Magnaudet}},\ and\
  \bibinfo {author} {\bibfnamefont {V.}~\bibnamefont {Pimienta}},\ }\href
  {https://doi.org/https://doi.org/10.1016/j.jcis.2022.05.154} {\bibfield
  {journal} {\bibinfo  {journal} {J. Colloid Interface Sci.}\ }\textbf
  {\bibinfo {volume} {625}},\ \bibinfo {pages} {990} (\bibinfo {year}
  {2022})}\BibitemShut {NoStop}%
\bibitem [{\citenamefont {Pimienta}\ \emph {et~al.}(2011)\citenamefont
  {Pimienta}, \citenamefont {Brost}, \citenamefont {Kovalchuk}, \citenamefont
  {Bresch},\ and\ \citenamefont
  {Steinbock}}]{https://doi.org/10.1002/anie.201104261}%
  \BibitemOpen
  \bibfield  {author} {\bibinfo {author} {\bibfnamefont {V.}~\bibnamefont
  {Pimienta}}, \bibinfo {author} {\bibfnamefont {M.}~\bibnamefont {Brost}},
  \bibinfo {author} {\bibfnamefont {N.}~\bibnamefont {Kovalchuk}}, \bibinfo
  {author} {\bibfnamefont {S.}~\bibnamefont {Bresch}},\ and\ \bibinfo {author}
  {\bibfnamefont {O.}~\bibnamefont {Steinbock}},\ }\href
  {https://doi.org/https://doi.org/10.1002/anie.201104261} {\bibfield
  {journal} {\bibinfo  {journal} {Angew. Chem. Int. Ed.}\ }\textbf {\bibinfo
  {volume} {50}},\ \bibinfo {pages} {10728} (\bibinfo {year}
  {2011})}\BibitemShut {NoStop}%
\bibitem [{\citenamefont {Banno}\ \emph {et~al.}(2016)\citenamefont {Banno},
  \citenamefont {Asami}, \citenamefont {Ueno}, \citenamefont {Kitahata},
  \citenamefont {Koyano}, \citenamefont {Asakura},\ and\ \citenamefont
  {Toyota}}]{banno2016deformable}%
  \BibitemOpen
  \bibfield  {author} {\bibinfo {author} {\bibfnamefont {T.}~\bibnamefont
  {Banno}}, \bibinfo {author} {\bibfnamefont {A.}~\bibnamefont {Asami}},
  \bibinfo {author} {\bibfnamefont {N.}~\bibnamefont {Ueno}}, \bibinfo {author}
  {\bibfnamefont {H.}~\bibnamefont {Kitahata}}, \bibinfo {author}
  {\bibfnamefont {Y.}~\bibnamefont {Koyano}}, \bibinfo {author} {\bibfnamefont
  {K.}~\bibnamefont {Asakura}},\ and\ \bibinfo {author} {\bibfnamefont
  {T.}~\bibnamefont {Toyota}},\ }\href@noop {} {\bibfield  {journal} {\bibinfo
  {journal} {Sci. Rep.}\ }\textbf {\bibinfo {volume} {6}},\ \bibinfo {pages}
  {31292} (\bibinfo {year} {2016})}\BibitemShut {NoStop}%
\bibitem [{\citenamefont {Sumino}\ \emph {et~al.}(2007)\citenamefont {Sumino},
  \citenamefont {Kitahata}, \citenamefont {Seto},\ and\ \citenamefont
  {Yoshikawa}}]{PhysRevE.76.055202}%
  \BibitemOpen
  \bibfield  {author} {\bibinfo {author} {\bibfnamefont {Y.}~\bibnamefont
  {Sumino}}, \bibinfo {author} {\bibfnamefont {H.}~\bibnamefont {Kitahata}},
  \bibinfo {author} {\bibfnamefont {H.}~\bibnamefont {Seto}},\ and\ \bibinfo
  {author} {\bibfnamefont {K.}~\bibnamefont {Yoshikawa}},\ }\href
  {https://doi.org/10.1103/PhysRevE.76.055202} {\bibfield  {journal} {\bibinfo
  {journal} {Phys. Rev. E}\ }\textbf {\bibinfo {volume} {76}},\ \bibinfo
  {pages} {055202} (\bibinfo {year} {2007})}\BibitemShut {NoStop}%
\bibitem [{\citenamefont {Dwivedi}\ \emph {et~al.}(2023)\citenamefont
  {Dwivedi}, \citenamefont {Shrivastava}, \citenamefont {Pillai}, \citenamefont
  {Tiwari},\ and\ \citenamefont {Mangal}}]{D3SM00228D}%
  \BibitemOpen
  \bibfield  {author} {\bibinfo {author} {\bibfnamefont {P.}~\bibnamefont
  {Dwivedi}}, \bibinfo {author} {\bibfnamefont {A.}~\bibnamefont
  {Shrivastava}}, \bibinfo {author} {\bibfnamefont {D.}~\bibnamefont {Pillai}},
  \bibinfo {author} {\bibfnamefont {N.}~\bibnamefont {Tiwari}},\ and\ \bibinfo
  {author} {\bibfnamefont {R.}~\bibnamefont {Mangal}},\ }\href
  {https://doi.org/10.1039/D3SM00228D} {\bibfield  {journal} {\bibinfo
  {journal} {Soft Matter}\ }\textbf {\bibinfo {volume} {19}},\ \bibinfo {pages}
  {3783} (\bibinfo {year} {2023})}\BibitemShut {NoStop}%
\bibitem [{\citenamefont {Sakamoto}\ \emph {et~al.}(2022)\citenamefont
  {Sakamoto}, \citenamefont {Izri}, \citenamefont {Shimamoto}, \citenamefont
  {Miyazaki},\ and\ \citenamefont {Maeda}}]{doi:10.1073/pnas.2121147119}%
  \BibitemOpen
  \bibfield  {author} {\bibinfo {author} {\bibfnamefont {R.}~\bibnamefont
  {Sakamoto}}, \bibinfo {author} {\bibfnamefont {Z.}~\bibnamefont {Izri}},
  \bibinfo {author} {\bibfnamefont {Y.}~\bibnamefont {Shimamoto}}, \bibinfo
  {author} {\bibfnamefont {M.}~\bibnamefont {Miyazaki}},\ and\ \bibinfo
  {author} {\bibfnamefont {Y.~T.}\ \bibnamefont {Maeda}},\ }\href
  {https://doi.org/10.1073/pnas.2121147119} {\bibfield  {journal} {\bibinfo
  {journal} {Proc. Natl. Acad. Sci.}\ }\textbf {\bibinfo {volume} {119}},\
  \bibinfo {pages} {e2121147119} (\bibinfo {year} {2022})}\BibitemShut
  {NoStop}%
\bibitem [{\citenamefont {Ebata}\ and\ \citenamefont
  {Sano}(2015)}]{ebata2015swimming}%
  \BibitemOpen
  \bibfield  {author} {\bibinfo {author} {\bibfnamefont {H.}~\bibnamefont
  {Ebata}}\ and\ \bibinfo {author} {\bibfnamefont {M.}~\bibnamefont {Sano}},\
  }\href@noop {} {\bibfield  {journal} {\bibinfo  {journal} {Sci. Rep.}\
  }\textbf {\bibinfo {volume} {5}},\ \bibinfo {pages} {8546} (\bibinfo {year}
  {2015})}\BibitemShut {NoStop}%
\bibitem [{\citenamefont {Thiele}\ \emph {et~al.}(2004)\citenamefont {Thiele},
  \citenamefont {John},\ and\ \citenamefont {B\"ar}}]{PhysRevLett.93.027802}%
  \BibitemOpen
  \bibfield  {author} {\bibinfo {author} {\bibfnamefont {U.}~\bibnamefont
  {Thiele}}, \bibinfo {author} {\bibfnamefont {K.}~\bibnamefont {John}},\ and\
  \bibinfo {author} {\bibfnamefont {M.}~\bibnamefont {B\"ar}},\ }\href
  {https://doi.org/10.1103/PhysRevLett.93.027802} {\bibfield  {journal}
  {\bibinfo  {journal} {Phys. Rev. Lett.}\ }\textbf {\bibinfo {volume} {93}},\
  \bibinfo {pages} {027802} (\bibinfo {year} {2004})}\BibitemShut {NoStop}%
\bibitem [{\citenamefont {Bates}\ \emph {et~al.}(2008)\citenamefont {Bates},
  \citenamefont {Stevens}, \citenamefont {Langford},\ and\ \citenamefont
  {Dickinson}}]{doi:10.1021/la800105h}%
  \BibitemOpen
  \bibfield  {author} {\bibinfo {author} {\bibfnamefont {C.~M.}\ \bibnamefont
  {Bates}}, \bibinfo {author} {\bibfnamefont {F.}~\bibnamefont {Stevens}},
  \bibinfo {author} {\bibfnamefont {S.~C.}\ \bibnamefont {Langford}},\ and\
  \bibinfo {author} {\bibfnamefont {J.~T.}\ \bibnamefont {Dickinson}},\ }\href
  {https://doi.org/10.1021/la800105h} {\bibfield  {journal} {\bibinfo
  {journal} {Langmuir}\ }\textbf {\bibinfo {volume} {24}},\ \bibinfo {pages}
  {7193} (\bibinfo {year} {2008})}\BibitemShut {NoStop}%
\bibitem [{\citenamefont {Ohta}\ and\ \citenamefont
  {Ohkuma}(2009)}]{PhysRevLett.102.154101}%
  \BibitemOpen
  \bibfield  {author} {\bibinfo {author} {\bibfnamefont {T.}~\bibnamefont
  {Ohta}}\ and\ \bibinfo {author} {\bibfnamefont {T.}~\bibnamefont {Ohkuma}},\
  }\href {https://doi.org/10.1103/PhysRevLett.102.154101} {\bibfield  {journal}
  {\bibinfo  {journal} {Phys. Rev. Lett.}\ }\textbf {\bibinfo {volume} {102}},\
  \bibinfo {pages} {154101} (\bibinfo {year} {2009})}\BibitemShut {NoStop}%
\bibitem [{\citenamefont {Tarama}\ and\ \citenamefont
  {Ohta}(2016)}]{Tarama_2016}%
  \BibitemOpen
  \bibfield  {author} {\bibinfo {author} {\bibfnamefont {M.}~\bibnamefont
  {Tarama}}\ and\ \bibinfo {author} {\bibfnamefont {T.}~\bibnamefont {Ohta}},\
  }\href {https://doi.org/10.1209/0295-5075/114/30002} {\bibfield  {journal}
  {\bibinfo  {journal} {Europhys. Lett.}\ }\textbf {\bibinfo {volume} {114}},\
  \bibinfo {pages} {30002} (\bibinfo {year} {2016})}\BibitemShut {NoStop}%
\bibitem [{\citenamefont {Ohta}(2017)}]{doi:10.7566/JPSJ.86.072001}%
  \BibitemOpen
  \bibfield  {author} {\bibinfo {author} {\bibfnamefont {T.}~\bibnamefont
  {Ohta}},\ }\href {https://doi.org/10.7566/JPSJ.86.072001} {\bibfield
  {journal} {\bibinfo  {journal} {J. Phys. Soc. Jpn.}\ }\textbf {\bibinfo
  {volume} {86}},\ \bibinfo {pages} {072001} (\bibinfo {year}
  {2017})}\BibitemShut {NoStop}%
\bibitem [{\citenamefont {Ohta}\ \emph {et~al.}(2016)\citenamefont {Ohta},
  \citenamefont {Tarama},\ and\ \citenamefont {Sano}}]{OhtaSano}%
  \BibitemOpen
  \bibfield  {author} {\bibinfo {author} {\bibfnamefont {T.}~\bibnamefont
  {Ohta}}, \bibinfo {author} {\bibfnamefont {M.}~\bibnamefont {Tarama}},\ and\
  \bibinfo {author} {\bibfnamefont {M.}~\bibnamefont {Sano}},\ }\href
  {https://doi.org/10.1016/j.physd.2015.10.007} {\bibfield  {journal} {\bibinfo
   {journal} {Physica D: Nonlinear Phenomena}\ }\textbf {\bibinfo {volume}
  {318-319}},\ \bibinfo {pages} {3} (\bibinfo {year} {2016})}\BibitemShut
  {NoStop}%
\bibitem [{\citenamefont {Ebata}\ \emph {et~al.}(2018)\citenamefont {Ebata},
  \citenamefont {Yamamoto}, \citenamefont {Tsuji}, \citenamefont {Sasaki},
  \citenamefont {Moriyama}, \citenamefont {Kuboki},\ and\ \citenamefont
  {Kidoaki}}]{EbataKidoaki}%
  \BibitemOpen
  \bibfield  {author} {\bibinfo {author} {\bibfnamefont {H.}~\bibnamefont
  {Ebata}}, \bibinfo {author} {\bibfnamefont {A.}~\bibnamefont {Yamamoto}},
  \bibinfo {author} {\bibfnamefont {Y.}~\bibnamefont {Tsuji}}, \bibinfo
  {author} {\bibfnamefont {S.}~\bibnamefont {Sasaki}}, \bibinfo {author}
  {\bibfnamefont {K.}~\bibnamefont {Moriyama}}, \bibinfo {author}
  {\bibfnamefont {T.}~\bibnamefont {Kuboki}},\ and\ \bibinfo {author}
  {\bibfnamefont {S.}~\bibnamefont {Kidoaki}},\ }\href
  {https://doi.org/10.1038/s41598-018-23540-x} {\bibfield  {journal} {\bibinfo
  {journal} {Sci. Rep.}\ }\textbf {\bibinfo {volume} {8}},\ \bibinfo {pages}
  {5153} (\bibinfo {year} {2018})}\BibitemShut {NoStop}%
\bibitem [{\citenamefont {L\"{o}ffler}\ \emph {et~al.}(2018)\citenamefont
  {L\"{o}ffler}, \citenamefont {Gorecki},\ and\ \citenamefont
  {Hanczyc}}]{10.1162/isal_a_00106}%
  \BibitemOpen
  \bibfield  {author} {\bibinfo {author} {\bibfnamefont {R.~J.}\ \bibnamefont
  {L\"{o}ffler}}, \bibinfo {author} {\bibfnamefont {J.}~\bibnamefont
  {Gorecki}},\ and\ \bibinfo {author} {\bibfnamefont {M.}~\bibnamefont
  {Hanczyc}},\ }\href {https://doi.org/10.1162/isal_a_00106} {\bibfield
  {journal} {\bibinfo  {journal} {Proceedings of the ALIFE 2018: The 2018
  Conference on Artificial Life. Tokyo, Japan}\ ,\ \bibinfo {pages} {574}}
  (\bibinfo {year} {2018})}\BibitemShut {NoStop}%
\bibitem [{\citenamefont {L\"{o}ffler}(2021)}]{Loeffler_Thesis}%
  \BibitemOpen
  \bibfield  {author} {\bibinfo {author} {\bibfnamefont {R.~J.}\ \bibnamefont
  {L\"{o}ffler}},\ }\emph {\bibinfo {title} {New materials for studies on
  nanostructures and spatio-temporal patterns self-organized by surface
  phenomena}},\ \href@noop {} {\bibinfo {type} {Phd thesis}},\ \bibinfo
  {school} {Institute of Physical Chemistry, Polish Academy of Sciences,
  https://ichf.edu.pl/files/tytuly/loffler-phd-thesis.pdf} (\bibinfo {year}
  {2021})\BibitemShut {NoStop}%
\bibitem [{\citenamefont {Schneider}\ \emph {et~al.}(2012)\citenamefont
  {Schneider}, \citenamefont {Rasband},\ and\ \citenamefont
  {Eliceiri}}]{ImageJ}%
  \BibitemOpen
  \bibfield  {author} {\bibinfo {author} {\bibfnamefont {C.~A.}\ \bibnamefont
  {Schneider}}, \bibinfo {author} {\bibfnamefont {W.~S.}\ \bibnamefont
  {Rasband}},\ and\ \bibinfo {author} {\bibfnamefont {K.~W.}\ \bibnamefont
  {Eliceiri}},\ }\href@noop {} {\bibfield  {journal} {\bibinfo  {journal} {Nat.
  Methods}\ }\textbf {\bibinfo {volume} {9}},\ \bibinfo {pages} {671} (\bibinfo
  {year} {2012})}\BibitemShut {NoStop}%
\bibitem [{\citenamefont {Fordham}\ and\ \citenamefont
  {Freeth}(1948)}]{doi:10.1098/rspa.1948.0063}%
  \BibitemOpen
  \bibfield  {author} {\bibinfo {author} {\bibfnamefont {S.}~\bibnamefont
  {Fordham}}\ and\ \bibinfo {author} {\bibfnamefont {F.~A.}\ \bibnamefont
  {Freeth}},\ }\href {https://doi.org/10.1098/rspa.1948.0063} {\bibfield
  {journal} {\bibinfo  {journal} {Proc. R. Soc. London}\ }\textbf {\bibinfo
  {volume} {194}},\ \bibinfo {pages} {1} (\bibinfo {year} {1948})}\BibitemShut
  {NoStop}%
\bibitem [{\citenamefont {Lenavetier}\ \emph {et~al.}(2024)\citenamefont
  {Lenavetier}, \citenamefont {Aud\'eoud}, \citenamefont {Berry}, \citenamefont
  {Gauthier}, \citenamefont {Poryles}, \citenamefont {Tr\'egou\"et},\ and\
  \citenamefont {Cantat}}]{PRL_Lenavetier}%
  \BibitemOpen
  \bibfield  {author} {\bibinfo {author} {\bibfnamefont {T.}~\bibnamefont
  {Lenavetier}}, \bibinfo {author} {\bibfnamefont {G.}~\bibnamefont
  {Aud\'eoud}}, \bibinfo {author} {\bibfnamefont {M.}~\bibnamefont {Berry}},
  \bibinfo {author} {\bibfnamefont {A.}~\bibnamefont {Gauthier}}, \bibinfo
  {author} {\bibfnamefont {R.}~\bibnamefont {Poryles}}, \bibinfo {author}
  {\bibfnamefont {C.}~\bibnamefont {Tr\'egou\"et}},\ and\ \bibinfo {author}
  {\bibfnamefont {I.}~\bibnamefont {Cantat}},\ }\href
  {https://doi.org/10.1103/PhysRevLett.132.054001} {\bibfield  {journal}
  {\bibinfo  {journal} {Phys. Rev. Lett.}\ }\textbf {\bibinfo {volume} {132}},\
  \bibinfo {pages} {054001} (\bibinfo {year} {2024})}\BibitemShut {NoStop}%
\bibitem [{\citenamefont {Kovalchuk}\ \emph {et~al.}(1999)\citenamefont
  {Kovalchuk}, \citenamefont {Kamusewitz}, \citenamefont {Vollhardt},\ and\
  \citenamefont {Kovalchuk}}]{Kovalchuk}%
  \BibitemOpen
  \bibfield  {author} {\bibinfo {author} {\bibfnamefont {V.~I.}\ \bibnamefont
  {Kovalchuk}}, \bibinfo {author} {\bibfnamefont {H.}~\bibnamefont
  {Kamusewitz}}, \bibinfo {author} {\bibfnamefont {D.}~\bibnamefont
  {Vollhardt}},\ and\ \bibinfo {author} {\bibfnamefont {N.~M.}\ \bibnamefont
  {Kovalchuk}},\ }\href {https://doi.org/10.1103/PhysRevE.60.2029} {\bibfield
  {journal} {\bibinfo  {journal} {Phys. Rev. E}\ }\textbf {\bibinfo {volume}
  {60}},\ \bibinfo {pages} {2029} (\bibinfo {year} {1999})}\BibitemShut
  {NoStop}%
\bibitem [{\citenamefont {de~Gennes}\ \emph {et~al.}(2004)\citenamefont
  {de~Gennes}, \citenamefont {Brochard-Wyart},\ and\ \citenamefont
  {Qu{\'e}r{\'e}}}]{Gennes2004CapillarityAW}%
  \BibitemOpen
  \bibfield  {author} {\bibinfo {author} {\bibfnamefont {P.~G.}\ \bibnamefont
  {de~Gennes}}, \bibinfo {author} {\bibfnamefont {F.}~\bibnamefont
  {Brochard-Wyart}},\ and\ \bibinfo {author} {\bibfnamefont {D.}~\bibnamefont
  {Qu{\'e}r{\'e}}},\ } {\emph
  {\bibinfo {title} {Capillarity and Wetting Phenomena: Drops, Bubbles, Pearls, Waves}}}\ (\bibinfo  {publisher} {Springer},\ \bibinfo
  {year} {2004})\BibitemShut {NoStop}%
\end{thebibliography}
\end{document}